\documentclass[12 pt]{article}

\usepackage{hyperref}
\usepackage[utf8]{inputenc}
\usepackage[a4paper, total={6in, 8in}, margin = 2cm]{geometry}
\usepackage[english]{babel}
\usepackage{authblk}
\usepackage{enumitem}
\usepackage{mathtools}
\usepackage{amssymb}
\usepackage{amsmath}
\usepackage{amsthm}
\usepackage{dsfont}
\usepackage{setspace}
\usepackage[authoryear]{natbib}
\usepackage{multirow,array}
\usepackage{sgamex}
\usepackage{bbm}
\usepackage{graphicx}
\usepackage{array, makecell}
\usepackage{float}
\usepackage[linesnumbered,boxruled,lined]{algorithm2e}
\usepackage{algpseudocode}
\usepackage{booktabs}

\usepackage[labelfont=sf,hypcap=false]{caption}

\usepackage{todonotes}

\usepackage{setspace}

\usepackage{xcolor}
\newcommand{\new}[1]{{\color{black}#1}}

\newcommand{\cxi}{{\mathfrak{c}}}
\newcommand{\mfI}{{\mathfrak{I}}}
\newcommand{\mfJ}{{\mathfrak{J}}}
\newcommand{\mfM}{{\mathfrak{M}}}

\newcommand{\co}{\text{CO}_2\text{e}}

\newcommand{\tT}{{t\in[0,T]}}

\newcommand{\N}{{\mathds{N}}}
\newcommand{\R}{\mathds{R}}
\newcommand{\1}{\mathds{1}}

\newcommand{\E}{\mathbb{E}}
\renewcommand{\P}{\mathbb{P}}

\newcommand{\V}{\mathcal{V}}

\newcommand{\T}{\mathcal{T}}
\newcommand{\F}{\mathcal{F}}

\newcommand{\btau}{{\boldsymbol{\tau}}}
\newcommand{\bx}{{\boldsymbol{x}}}
\newcommand{\bi}{{\boldsymbol{i}}}
\newcommand{\bX}{{\boldsymbol{X}}}

\newcommand{\ind}{\mathds{1}}

\newcommand{\vx}{\partial_x U^\star}

\newcommand{\vs}{\partial_s U^\star}
\newcommand{\vstwo}{\partial_{ss} U^\star}
\newcommand{\vt}{\partial_{t} U^\star}

\renewcommand{\star}{{}}

\begin{document}

\title{Nash Equilibria in Greenhouse Gas Offset Credit Markets}
\author{
Liam Welsh\footnote{\href{mailto:liam.welsh@mail.utoronto.ca}{liam.welsh@mail.utoronto.ca}}
\\ Department of Statistical Sciences,\\ University of Toronto
\\
\and
Sebastian Jaimungal\footnote{\href{mailto:sebastian.jaimungal@utoronto.ca}{sebastian.jaimungal@utoronto.ca}}
\\
Department of Statistical Sciences,
\\
University of Toronto \\
\&
\\
Oxford-Man Institute for Quantitative Finance, \\
University of Oxford 
}
\bigbreak
\bigbreak
\maketitle

\begin{abstract}

\new{One approach to reducing greenhouse gas (GHG) emissions} is to incentivize carbon capturing and carbon reducing projects while simultaneously penalising excess GHG output. \new{In this work}, we present a novel market framework and characterise the optimal behaviour of GHG \new{offset credit (OC)} market participants in both single-player and two-player settings. The single player setting is posed as an optimal stopping and control problem, while the two-player setting is posed as optimal stopping and mixed-Nash equilibria problem. We demonstrate the importance of acting optimally using numerical solutions and Monte Carlo simulations and explore the differences between the homogeneous and heterogeneous players. \new{In both settings, we find that market participants  benefit from optimal OC trading and OC generation.}

\end{abstract}

\newpage

\section{Introduction}\label{sec:intro}

Severe impacts of man-made climate change are being experienced worldwide, caused by pollution and excess emissions since the industrial revolution. In an attempt to limit the damage of climate change, nations around the world have been organising and establishing international treaties, including the Kyoto Protocol in 1997, the Paris Agreement in 2015, and the Glasgow Climate Pact in 2021. These treaties aim to limit greenhouse gas (GHG) pollution, increase financing opportunities for developing nations to adopt climate friendly policy, and spur the development of green technologies. One tool that nations can use to curb GHG emissions is to implement carbon taxes or establish emissions markets. This work focuses on the latter, specifically the Canadian GHG Offset Credit (OC) market where market participants can earn OCs by implementing projects that reduce or capture GHG emissions. These OCs may then be used to offset excess emissions or may be sold to other firms.

Given the pressing nature of climate change and the popularity of climate finance, the resulting extant literature is growing, particularly for climate, energy, and emissions based financial derivatives.
In many countries, including Canada~(\cite{canadaFederalCarbon,canadaPriceCarbon}), impose their own carbon prices internally. While many regions are pricing carbon emissions locally, there lacks an agreed upon global floor price for carbon which could further incentivize emission reductions~(\cite{globalPriceIMF}). In fact, there are over sixty carbon pricing instruments worldwide to control carbon emissions, including emission markets and carbon taxes, covering $21\%$ of GHG emissions~(\cite{santikarn2021state}). However, only $4\%$ are valuing carbon emissions at a level inline with the Paris Agreement to meet the $2^{\circ}$C limit.

GHG OC markets can often be structured as carbon cap-\&-trade (C\&T) markets, of which there is considerable literature.~\cite{seifert2008dynamic} describe firm behaviour as the solution to an optimal control problem in a single period C\&T market, and further solve for the carbon allowance price process. This is extended by~\cite{hitzemann2018equilibrium} to allow for a multi-period model.~\cite{howison2012risk} presents a risk-neutral pricing framework for emissions markets derivatives, while~\cite{carmona2009optimal} analyses a player's optimal behaviour in carbon markets and further scrutinise the potential pitfalls of emission markets. More recent work by~\cite{carmona2022mean} investigate how carbon taxes influence energy producers' production methods in a mean-field setting. The authors incorporate social external negativity of carbon emissions in their model to achieve both a Nash equilibrium and what they deem to be a \textit{social optimum}. The notion of utilising social impact with regards to climate finance is not uncommon. In particular,~\cite{kiesel2022prolegomenon} manages climate change risk via an uncertainty-based framework, such that the authors claim capital-based risk measures are unable to properly capture the deep uncertainties induced by climate change.~\cite{kenyon2022carbon} introduce a \textit{carbon equivalence principle} to allow for better market alignment for $\co$ based derivatives for improved coherence in emissions markets. In green derivatives research,~\cite{amundsen2006price} analyses price volatility models of renewable energy certificates (RECs) in the context of energy. REC markets and green certificate markets are closely related alternatives to carbon emissions markets.~\cite{shrivatsSREC} characterises solar REC markets, which is then extended to a mean-field game setting in~\cite{Shrivats2022_mean} and \cite{firoozi2021principal}.~\cite{hustveit2017tradable} investigates green certificate markets from an investors perspective, and find that with increased renewable energy penetration, the value of green certificates will decline to zero despite their highly sensitive nature to energy generation and consumption.

Here, we develop a novel framework that characterises player behaviour in a multi-player GHG OC market in the form of a bi-matrix game. For our work, we employ techniques from stochastic control theory for optimal stopping~(\cite{pham2009continuous}) and game theory~(\cite{fudenberg1991game, osborne1994game,osborne2004introduction}) to characterise optimality in player behaviour. When there are many players in a market, mean-field game (MFG) (see~\cite{carmona2018probabilistic,huang2006large,huang2007large,lasry2007MFG}) may be used to study approximations to Nash equilibria. Canada's emissions market, however, is not highly populated~(\cite{sadikman2022evolution}), thus we do not proceed along a MFG approach. Instead, we analyse a GHG OC market using traditional game theoretic approaches for two cases: single-player and two-player. In the single-player setting, the player has {the ability to choose when to} generate OCs through project investment. This leads to a combined sequence of stopping and control problems and we solve the resulting quasi-variational inequality (QVI) using an implicit-explicit finite-difference (FD) scheme. As OCs must converge to their marginal price at the compliance date, we model OC price dynamics with a Brownian bridge. In the two-player setting, both players can choose to trade or generate OCs, therefore, the QVI is replaced with a bi-matrix game where each player's decision has an impact on the other's value. The induced Nash equilibrium from the bi-matrix game provides insight as to how players behave when presented with homogeneous and heterogeneous project opportunities. The approach we take can be extended to a multiple player setting, however, there are computational issues that need addressing to make it a viable approach. We also present simulations for both the single-player and two-player markets, and compared against na\"{i}ve strategies demonstrating the importance of properly modelling players' optimal behaviour. {\cite{liu2023purchase}, the closest work to ours, formulates a single player optimal stopping problem to analyse when a firm should purchase a carbon emission right for carbon asset management -- akin to an American option. Our work differs in multiple dimensions though, including the motivating setting, the single and multi-player formulation (as we have sequences of combined stopping and control problems), and the numerical experiments that we carry out.} 

The remainder of this paper is organised as follows. Necessary background information on the GHG market in Canada and motivation is provided in Section~\ref{sec:background}. Section~\ref{sec:market} introduces the GHG OC market structure used throughout this paper. Section~\ref{sec:onePlayer} presents the mathematical model used in a single-player setting, with a numerical implementation and simulation results in Sections~\ref{sec:numImpOne} and~\ref{sec:resOne}, respectively. In Section~\ref{sec:twoPlayer}, the two-player version of the market is introduced, and the bi-matrix game is constructed. Section~\ref{sec:numImpTwo} and~\ref{sec:resTwo} contain the numerical implementation and simulated results for the two-player setting, respectively. The two-player model is then extended to a multi-period setting in Section~\ref{sec:multi_period}. Concluding remarks and potential avenues of future research are provided in Section~\ref{sec:conc}.

\section{Background and Motivation}\label{sec:background}

In Canada, GHG emissions (also referred to as carbon emissions) are regulated at both the federal and provincial level. Provincial governments are allowed to implement their own GHG emissions framework and market (e.g.~a carbon price) with the caveat that it meets the minimum federal pricing standards and GHG reduction targets~(\cite{sadikman2022evolution}). The Canadian federal government has recently developed a new GHG OC system at the federal level aimed at expanding project opportunities and streamlining the process~(\citep{canadaGHGOC,sadikman2022evolution}). This new system contains three main components: a new regulations framework for OC generation; updated development protocols for methods to quantify GHG reductions across various sectors; and an OC tracking and project registration system. The updated framework is available for both regulated and unregulated firms. Any firm may participate in the voluntary market, but regulated firms must take part in the compliance market. We summarise the Canadian market structure below (see~\cite{canadaGHGOC, canadaGazetta_Outline, canadaFederalCarbon, canadaPriceCarbon, sadikman2022evolution}).

In the compliance market, firms are provided with an emissions limit in metric tonnes (Mt) of carbon dioxide equivalent ($\co$). A $\co$ is a predefined and common unit used for to measure GHG emissions based on their warming potential, relative to CO$_2$. For example, one tonne of methane has a global warming potential 25 times higher than that of CO$_2$, and hence one tonne of methane is equivalent to 25 tonnes of CO$_2$. Firms that are under their emissions limit receive surplus OCs equal to to the difference of their limit to their output. Firms that exceed their emissions limit are penalised and can use acquired OCs to reduce or eliminate their penalty. Penalised firms must either pay a fine or submit a combination of surplus and non-surplus OCs. Non-surplus OCs, which are federally or provincially approved and regulated, can be generated by the firm itself or purchased from other firms. In 2022, the federal government set the excess GHG emissions penalty to \$50 per Mt$\co$, and is increasing the penalty each year a linear amount each year until 2030 when the penalty will reach \$170 per Mt$\co$ as a method to financially motivate firms to generate OCs and participate in the GHG OC market.

OCs are certified financial derivatives distributed by a regulatory body that can be used by regulated firms for compliance purposes. OCs are measured in tonnes of CO$_2$e, such that one OC represents the removal (or reduction) of one tonne of CO$_2$e from the atmosphere or production process. For these OCs, GHGs are converted using their warming potential back into CO$_2$ to allow for standardisation across OC markets. The purpose of GHG OC markets is to encourage the development of GHG capturing and/or GHG reducing projects by GHG emitting firms, to slow the destruction of climate change. OCs are generated via approval from a governing body by reducing GHG emissions or by implementing and registering a GHG emissions reducing or capturing project. These projects typically have investment costs associated. Project examples include, but are not limited to, tree planting, capturing methane generated from landfills or livestock, composting, and updating machinery to more efficient and less polluting models. OCs can be invalidated after their creation if it was found that the OCs were issued under incorrect information. The regulating entity typically requires that these be replaced, thus an additional investment must be made by the firm in order to comply. Hence, an OC represents a verifiable receipt that either GHGs emissions have been reduced or GHGs have been captured from the atmosphere. On a compliance date, firms that exceed their emissions limit over the corresponding compliance period submit acquired OCs to the regulatory body to mitigate their financial penalty. OCs that are generated in the voluntary market cannot be used for regulatory purposes. As such, OCs that are generated in the compliance market typically trade at a premium over their voluntary market counterparts. Compliance market OCs remain viable for regulatory purposes for eight calendar years from the date of their creation. 

Both regulated and unregulated firms can generate and trade OCs. While regulated firms that take part in the compulsory market must abide by certain regulations or else face a penalty, unregulated firms can participate in the voluntary market. The motivation for unregulated firms to participate (and potentially invest in OC generation) can come from, but is not limited to, internal emission reduction commitments, ESG and other altruistic goals, or customer and investor pressure. {In 2023, the federal government began accepting and registering GHG OC generating projects. Due to the recency of this market's inception, there is currently limited data available, and as of February 2024, only one project has been registered and verified~\citep{fedTracking}. Provincial markets, however, have been established for many years, including British Columbia's (BCs) and Qu\'{e}bec's. Between the years of 2010 and 2022 in BC, 8.7 million tonnes of $\co$ were offset from the public sector by OCs generated within the province through registered and verified OC projects~\citep{offsetBC}, while in Qu\'{e}bec, over 1.5 million OCs were issued between January 2014 and March 2024 for projects including landfill methane reclamation/destruction and halocarbon destruction~\citep{offsetQUE}.}

The voluntary market in Canada is gaining traction, due to carbon neutrality goals set by many major banks, energy and natural resource firms, and public institutions. Hence, demand for OCs in the voluntary market is increasing. OCs that are generated in the voluntary market typically trade at a discount to those traded in the compliance market as the validating bodies may not meet the rigorous standards of the provincial or federal agencies. As such, not all voluntarily produced OCs may eligible to be used for compliance purposes. Unregulated firms may generate OCs that can be used for compliance purposes, assuming project and validating requirements were met. We analyse regulated firms participating in the compliance market. This choice is made for two reasons: (1) the goals of unregulated firms greatly vary within the voluntary market and are dependent on each firms reason for participation; and (2) due to the possibility of financial penalties, regulated firms are incentivized to take part in this market. These firms have the choice between investing in GHG reducing or capturing projects and trading for OCs on the market, inducing a game structure to the market model.

\section{Offset Credit Market Model}\label{sec:market}

We assume the following structure for our GHG OC market model. A firm (i.e.~player) that participates in the market may be regulated or unregulated. Regulated firms (i.e.~firms with emission thresholds) are required to participate in this market. This type of firm has a maximum threshold of emissions that if it exceeds, the firm  faces a financial penalty. In our setting, we assume that a regulated firm has an emission limit in units of Mt$\co$ at the end of a regulatory period at time $T$. Excess emissions over this limit, denoted by $R$, must be covered using OCs. Any OC shortfall from $R$ (and above the emissions limit) is penalised. We assume $R$ to be deterministic and exogenously determined. Throughout this work, we refer to $R$ as the OC requirement, as a player requires a terminal inventory of $R$ OCs to completely eliminate their penalty. The penalty structure implemented in Canada is linear and the penalty value per emission excess is growing each year~\citep{sadikman2022evolution}. Thus, we impose a penalty structure at the end of a compliance period of
\begin{equation}\label{eqn:penalty}
    G(x) := -p \,(R - x)_+\;,
\end{equation}
where $y_+ := \max(y,0)$ and $x$ represents a firm's OC inventory. It is possible that the terminal penalty can take a variety of forms depending on the market regulations. In the terminal penalty above, a firm is penalised a fixed value $p$ per unit of OC shortfall (i.e.~a linear penalty structure). Unregulated firms do not have to meet some minimum threshold of OC inventory. As such, the terminal value can take a more diverse array of possibilities that may incorporate social benefits or public relations metrics. The structure of the terminal value for an unregulated firm can vary largely depending on a firms goals, products, and/or clientele. Alternatively, the market can be structured such that it does not penalise inaction, but instead rewards participation. Using this style of terminal preferences can apply to both regulated and unregulated firms. A (monetary) reward can be given to firms for each OC they hold at the end of some period, either up to some maximum threshold or a diminishing value structure can be employed. The flexibility of this choice can allow market makers and legislators to analyse the potential impact on firms' behaviours.

A firm's OCs may either be generated by approval from a regulatory body through project investment or can be bought and/or sold on the GHG OC market. We assume that at all instances in time $t\in[0,T]$, a firm may generate OCs. In particular, the firm aims to decide on the rate of OC investment and the sequence of times at which it will invest in an eligible project. {A firm may also trade OCs throughout the entire period, and the decision to generate does not impact their ability to trade}. The single and two-player versions of the problem are analysed in Sections~\ref{sec:onePlayer} and~\ref{sec:twoPlayer}, respectively. 

We work on a completed and filtered probability space $(\Omega, \F, (\F_t)_\tT,\P)$, on which the OC spot price is denoted $S =\left(S_t\right)_\tT$, and $(\F_t)_\tT$ is the natural filtration generated by $S$.
The firm's OC inventory process is defined as $X=\left(X_t\right)_{t\in[0,T]}$.  A firm can trade at rate $\nu = \left(\nu_t\right)_{t\in[0,T]}$, such that $\nu\in\mathcal{V}$ where $\mathcal{V}$ is the set of admissible trading rates consisting of $\F$-predictable process such that $\E[\int_0^T \nu_t^2\,dt]<\infty$. A positive (negative) rate corresponds to buying (selling) OCs.  From the purchase and sale of OCs the firm obtains a total reward of $-\int_0^T S_t\,\nu_t\,dt$. We further apply a trading friction in the form of a stylised transaction cost of $\frac{\kappa}{2}\int_0^T \nu_t^2\,dt$. This results in firms mitigating the speed at which they trade, and it may represent a combination of real and fictitious costs. While we specify a quadratic stylised transaction cost, any convex function of $\nu_t$ can be used to impose such a constraint.
The firm's OC inventory between times at which it invests satisfies the ODE $dX_t = \nu_t\, dt$. 

We assume that a firm has a fixed generation capacity of $\xi$, such that if a firm chooses to generate at time $\tau \in[0,T]$, their inventory jumps by $\xi$, so that $X_\tau  = X_{\tau ^-} + \xi$, where $X_{t^-}:=\lim_{s\uparrow t}X_s$.  The cost of generating these OCs is denoted $\cxi$. As the generation of OCs increases the possible trading volume, it is natural to include a price impact to the OC spot price in the event a firm generates. This price impact parameter is denoted $\eta$. Thus, if a firm generates $\xi$-many OCs at time $t\in[0,T]$, then the OC spot price jumps such that $S_\tau = S_{\tau ^-} - \eta\,\xi$. In the sequel, we denote the set of $\F$-stopping times by $\mathcal{T}$.

Next, we specify the dynamics of the OC spot price. In particular, we choose to model the price as a Brownian bridge, such that at $T$ the spot price equals the marginal price of an OC induced by the penalty given in \eqref{eqn:penalty}. Therefore, $S$ satisfies the SDE
\begin{equation}\label{eqn:S-SDE}
    dS_t = \frac{p-S_t}{T-t}\,dt + \sigma \, dW_t\;,    
\end{equation}
where $\sigma$ is the volatility of the OC spot price and $W=\left(W_t\right)_{t\in[0,T]}$ is a standard $\P$-Brownian motion. 

The choice of dynamics for the OC price forces the value of the OC spot price at time $t=T$ to be equal to the penalty value $p$ and the OC price also converges to the penalty value $p$ as $t\rightarrow T$, thus avoiding any potential arbitrage opportunity that firms may take advantage of. For instance, under other OC price dynamics (such as geometric Brownian motion), the OC spot price an instant prior to the terminal time could exceed the penalty value ($S_{T^-} > p$), such that a player would sell their OC inventory as they can generate more value from selling their inventory and accepting the penalty than by using their inventory for regulatory purposes. In a multiplayer scenario, players share a common noise due to the stochasticity of the OC price. We assume that no trading or generating can take place at the terminal time $t=T$, hence any OCs that are to be used for regulatory purposes must be procured prior to the regulatory date. Moreover, any excess OCs after submitting to the required amount to regulator expire worthless.

\section{Single-Player Setting}\label{sec:onePlayer}

In this section, we analyse how a single player participating in this market will optimally behave. This player has the capacity to generate $\xi$-many OCs at a cost of $\cxi$. The player's terminal value $V(T,x,s)$, where $V$ is the player's value, is determined by the terminal penalty function, $G(x) = -p\,(R-x)_+$.  The player optimises over trading speed $\nu\in\V$ and over an increasing sequence of $\F$-stopping times $\btau=(\tau_i)_{i\in\N}$, i.e., $\tau_i\in\T$, for all $i\in\N$ and $\tau_1<\tau_2<\dots$. The sequence of stopping times represents the times at which the player invests in a green project.
The player's performance criterion may be stated as
\begin{equation}\label{eq:oneP_tradeCost}
\E \bigg[G(X_T) - \int_0^{T} S_u\,\nu_u \,du - \frac{\kappa}{2}\int_0^{T} \nu^2_u\, du - \sum_{i\in\N} \ind_{\tau_i\le T}\,\cxi \bigg]\;.
\end{equation}
The first term represents the regulatory penalty, the second term  represents the trading costs, the third term represents the stylised transaction cost, and the fourth term represents the total generation costs.

As per the dynamic programming principle (DPP), to optimise the criterion over all $\nu\in\V$ and $\btau\in\T$, we introduce the value of an arbitrary strategy starting at an arbitrary point in time $t\in[0,T]$ at an arbitrary point in state space $X_{t^-}=x, S_{t^-}=s$, which we denote by 
\begin{equation}
    J^{\nu,\btau}(t,x,s)
    :=\E_{t,x,s}\left[
    G(X_T) - \int_t^{T} S_u\,\nu_u \,du - \frac{\kappa}{2}\int_t^{T} \nu^2_u\, du - \sum_{i\in\N} \ind_{t\le\tau_i\le T}\,\cxi
    \right]\,.
\end{equation}
Here,  $\E_{t,x,s}[\cdot]$ denotes conditional expectation given that  $X_{t^-}=x, S_{t^-}=s$.
Let $k:=\min\{i: \tau_i>t\}$, then, by iterated expectations, we may also write
\begin{equation}
\begin{split}
    J^{\nu,\btau}(t,x,s)
    :=&\;\E_{t,x,s}\Bigg[
    - \int_t^{\tau_k\wedge T} S_u\,\nu_u \,du - \frac{\kappa}{2}\int_t^{\tau_k\wedge T}
    \nu^2_u\, du 
    \\
    &\hspace*{4em}
    + \left( J^{\nu,\btau}
    \left(
    \tau_k, X_{\tau_k^-}+\xi, S_{\tau_k^-}-\eta\,\xi
    \right)
    -\cxi
    \right)
    \ind_{\{\tau_k\le T\}}
    \Bigg]\,.
\end{split}
\end{equation}

Next, define the value function as $V(t,x,s)=\sup_{\nu\in\V,\;\btau\in\T} J^{\nu,\btau}(t,x,s)$.
Applying the DPP for stopping and control problems (see~\cite{pham2009continuous}), the value function is the unique viscosity solution of the QVI
\begin{equation}\label{eq:single_QVI}
\begin{split}
       \max\Bigg\{ 
       &
   \partial_t V(t,x,s) +
   \sup_{v\in\R} \left( -v \,s - \frac{\kappa}{2}\,v^2  + v\, \partial_xV(t,x,s) + \frac{p-s}{T-t} \,\partial_sV(t,x,s) + \frac{\sigma^2}{2}\,\partial_{ss}V(t,x,s)\right)
   ;
   \\
   &
   \quad
   \big(V(t,x+\xi,s-\eta\,\xi)-\cxi\big)
   -V(t,x,s)
   \Bigg\} = 0\;,
\end{split}
\end{equation}
subject to the terminal condition $V(T,x,s) = G(x)$.

The continuation component of the QVI results in the player's optimal trading behaviour, while the second component of the QVI represents the change in the value function when they generate OCs which induces a change in the OC inventory and the asset price, as well as incurring a cost.

From the form above, we may solve for the optimal trade rate $v$ in feedback form as
\begin{equation}
v^*(t,x,s) = \frac{1}{\kappa}\left(\partial_x V(t,x,s) - s\right)\,.
\label{eqn:optimal-trading-rate-single-player}
\end{equation}
Substituting $v^*$ into the QVI \eqref{eq:single_QVI} results in the value function satisfying
\begin{equation}\label{eq:single_QVI_opt}
\begin{split}
       \max\Bigg\{ 
       &
   \partial_t V(t,x,s) +\tfrac{1}{2\kappa}
   \big(\partial_x V(t,x,s) - s\big)^2 +  \frac{p-s}{T-t} \,\partial_s V(t,x,s) + \frac{\sigma^2}{2}\,\partial_{ss} V(t,x,s)
   \;;
   \\
   &
   \quad
   \big(V(t,x+\xi,s-\eta\,\xi)-\cxi\big)
   -V(t,x,s)
   \Bigg\} = 0\;,
\end{split}
\end{equation}
subject to the terminal condition $V(T,x,s)=G(x)$. This QVI does not admit a closed form solution. Instead, we develop a numerical scheme for approximating the optimal solution in the following section.

\subsection{Numerical Implementation}\label{sec:numImpOne}

Numerically approximating the optimal solution of~\eqref{eq:single_QVI_opt} requires the estimation of two PDE solutions, which are derived from the value of a player trading and the value of a player generating. To develop the appropriate PDEs, we discretise time and write the time grid as $\{0,\Delta t, 2\Delta t, \dots, N\Delta t=T\}$ and denote $t_i:=i\,\Delta t$.

First, we denote the value in the continuation region (i.e., trading) as $U^\star$. From the dynamic programming principal, for $t\in[t_{k-1},t_{k}]$, $U^\star$ is given by
\begin{equation}\label{eq:oneP_noGenCost}
    U^\star(t,x,s) = \sup_{\nu\in\mathcal{V}}\;\E_{t,x,s}\left[V(t_k,X_{t_k},S_{t_k}) - \int_t^{t_k} S_u\,\nu_u\, du - \frac{\kappa}{2}\int_t^{t_k} \nu^2_u\, du \right]\;.
\end{equation}
The optimised profit functional~\eqref{eq:oneP_noGenCost} elicits the same form of the continuation component in~\eqref{eq:single_QVI_opt}. Through Feynman-Kac, $U^\star$ satisfies the PDE
\begin{equation}\label{eq:oneP_U_HJB}
    0 = \vt(t,x,s) +\frac{1}{2\kappa}\left(\vx(t,x,s) - s\right)^2 +  \frac{p-s}{T-t} \,\vs(t,x,s) + \frac{\sigma^2}{2}\,\vstwo(t,x,s)\;, 
\end{equation}
for all $t\in[t_{k-1}, t_k]$, and subject to terminal condition $U^\star(t_{k},x,s) = V(t_k,x,s)$.

The player chooses {whether or not to generate OCs}, hence, the value function at $t=t_{k-1}$ is given by
\begin{equation}\label{eq:comp_Opt}
    V(t_{k-1},x,s) = \max\left\{U^\star(t_{k-1},x,s),~  {\left(U^\star(t_{k-1},x+\xi,s-\eta\,\xi)-\cxi\right)} \right\}
\end{equation}
with terminal condition $V(T,x,s) = G(x)$.

To estimate the solution of~\eqref{eq:oneP_U_HJB} and solve~\eqref{eq:comp_Opt}, we implement an FD scheme that is implicit with respect to the OC price $s$ and explicit with respect to player's inventory $x$. This structure is chosen as it avoids numerical instabilities that arise when $t$ is close to the terminal time $T$ resulting from the Brownian bridge dynamics. We construct spatial grid $x_1, \dots, x_I$ with $x_{i+1}= x_i + \Delta x$ and offset price grid $s_1, \dots, s_J$ with $s_{j+1}= s_j + \Delta s$, and let $V_{k,i,j} = V(t_k,  x_i , s_j)$. Denote $\mfI:=\{1,\dots,I\}$ and $\mfJ:=\{1,\dots,J\}$. Further, we assume that the second partial derivatives in offset credit price vanish at the boundary points of the spatial grid. Thus, for $\mfI\setminus\{1,I\}$ we are required to solve the matrix system
\begin{equation} \label{eq:implicit_oneTrade}
    \begin{bmatrix} 
    1 & -2 & 1 & 0 & \dots \\
    a_2 & b_2 & c_2 & 0 & \dots \\
    0 & a_3 & b_3 & c_3 & 0 & \dots \\
    \vdots & \ddots & \ddots &\ddots & \ddots & \ddots \\ \\
    0 & \dots & & & a_{J-1} & b_{J-1} & c_{J-1} \\
    0 & \dots & & 0 & 1 & -2 & 1
    \end{bmatrix}
    \begin{bmatrix}
           U^{\star}_{k-1,i,1} \\
           U^{\star}_{k-1,i,2} \\
           U^{\star}_{k-1,i,3} \\
           \vdots \\ \\
           U^{\star}_{k-1,i,J-1} \\
           U^{\star}_{k-1,i,J}
         \end{bmatrix}
    =
    \begin{bmatrix}
           0 \\
           H_{k,i,2} \\
           H_{k,i,3} \\
           \vdots \\ \\
           H_{k,i,J-1} \\
           0
         \end{bmatrix}
\end{equation}
for the vector $U^{\star}_{k-1,i,\cdot}$ where
\begin{align}
    H_{k,i,j} &= {V_{k,i,j}} + \frac{\Delta t}{2\kappa}\left(\frac{V_{k,i+1,j}-V_{k,i-1,j}}{2\,\Delta x} - s_j \right)^2,
    \\
    a_j &= \Delta t\left(\frac{p-s_j}{2\,\Delta s\,(T-t_{k-1})} - \frac{\sigma^2}{2\,(\Delta s)^2}\right),
    \\
    b_j &= 1 + \Delta t\,\frac{\sigma^2}{(\Delta s)^2}\,,
    \qquad \text{and}
    \\
    c_j &= -\Delta t\left(\frac{p-s_j}{2\,\Delta s\,(T-t_{k-1})} + \frac{\sigma^2}{2\,(\Delta s)^2}\right)\,.
\end{align}

The values of $H$, $a$, $b$, and $c$ are obtained using centralised FD approximations to the PDE in~\eqref{eq:oneP_U_HJB} with an implicit relation in $\vs$ and $\vstwo$ while maintaining an explicit relation for $\vx$. The derivation of these values is provided in Appendix~\ref{app:coefs}.

Using the matrix system~\eqref{eq:implicit_oneTrade}, the algorithm for the one-player model is provided in Algorithm \ref{algo:single-player}.

\begin{algorithm}[H]
	\caption{Single-player value function and optimal decision boundary}\label{algo:single-player}
    \footnotesize
	\KwIn{t,s,x grid}
    Set terminal values $V_{N,i,j}=G(x_i)$\;
    \For{$k = N, N-2, \ldots,1$}{
        Solve \eqref{eq:implicit_oneTrade} for $U^{\star}_{k-1,i,\cdot}$ for each $i\in\mfI\setminus\{1,I\}$\;
        Use interpolation to compute $U^{\star}_{k-1,1,\cdot}$ and $U^{\star}_{k-1,I,\cdot}$\;
        Set $V_{k-1,i,j} = \max \left\{U^{\star}(t_{k-1},x_i,s_j),~{\left(U^\star(t_{k-1},x+\xi,s-\eta\,\xi)-\cxi\right)} \right\}$ for all $i\in\mfI,j\in\mfJ$\;
        Store the decision region $D_{k-1,i,j}:=\mathds{1}_{\{V_{k-1,i,j}\ge U^{\star}(t_{k-1},x_i,s_j)\}}$for all $i\in\mfI,j\in\mfJ$\;
        Store the optimal trading rate $v^*_{k-1,i,i}$ using the FD approximation of \eqref{eqn:optimal-trading-rate-single-player} for all $i\in\mfI,j\in\mfJ$\;
    }
    \KwOut{$V$, $\nu^*$, and $D$}
\end{algorithm}

\subsection{Results}\label{sec:resOne}

We implement the above algorithm in Python and generate $5,000$ OC price sample paths to analyse how a regulated player's optimal strategy behaves. We set the marginal cost of OC generation to be equal to the terminal penalty for all experiments. Specifically, in this single-player environment we take $\xi = 0.1$ OCs, $\cxi = \$0.25$, $R = 5$, and the penalty $p=\$2.50$ per OC below the requirement $R$. The model parameters are provided in Table~\ref{tab:base_params}. We use 100 time steps between the initial time $t=0$ and the terminal time $t = T$ and set the compliance period to be one month, $T = 1/12$ years. Unless otherwise specified, the units of time are in years. For the simulated OC price, we set a reflective boundary at zero. The volatility $\sigma$ of the OC price  is set to $0.5$.
\begin{table}[H]
\centering
\begin{tabular}{ ccccccccc } 
\toprule\toprule
 T (years) & $\sigma$ & $\kappa$ & $\eta$ & $\xi$ & $\cxi$ & $S_0$ & $R$ & Penalty \\ 
 \midrule
 1/12 & 0.5 & 0.03 & 0.05 & 0.1 & 0.25 & 2.5 & 5 & 2.5 \\ 
 \bottomrule\bottomrule
\end{tabular}
\caption{Simulation parameters for single-player market.}
\label{tab:base_params}
\end{table}

Three na\"{i}ve strategies are developed for comparative purposes: (1) trading at a constant rate $\tilde v$ such that the OC requirement is met at $t = T$; (2) trading at the constant rate $\tilde v$ until $t = T/2$ followed by successive generation until the requirement is met; and (3) successive generation starting at $t=0$ until the requirement is met. The same random seed is used to generate all scenarios for all strategies to allow for direct comparisons.
\begin{figure}[H]
\centering
\includegraphics[width=0.85\textwidth]{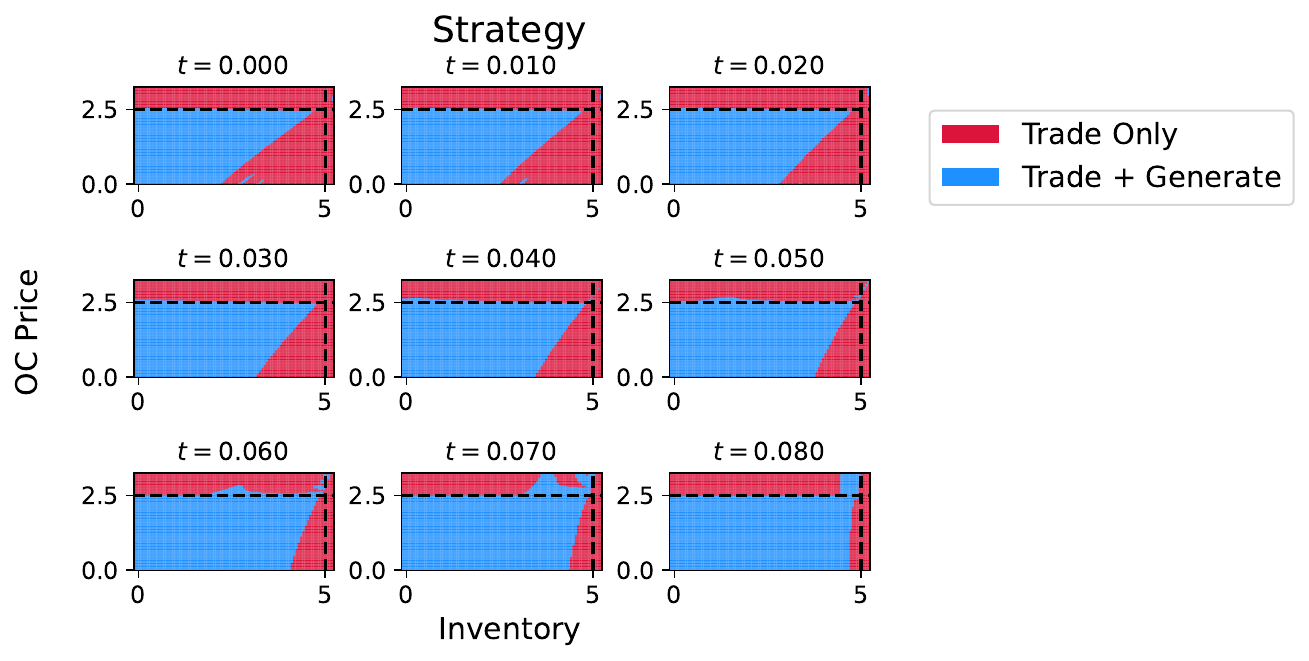}
\caption{Trade-Generation regions at multiple time points. Generation region is in white. Dashed lines indicate terminal OC requirement (vertical) and penalty value (horizontal).}
\label{fig:boundary_single_base}
\end{figure}

\begin{figure}[H]
\centering
\includegraphics[width=0.85\textwidth]{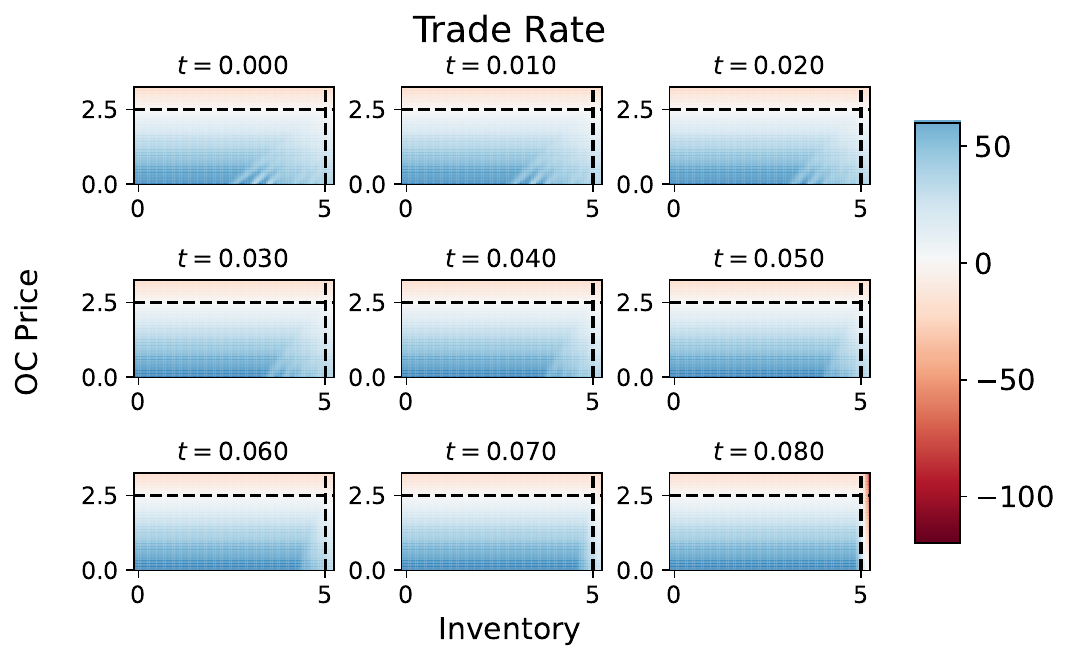}
\caption{{Trade rates at multiple time points. Dashed lines indicate terminal OC requirement (vertical) and penalty value (horizontal).}}
\label{fig:boundary_single_base_trade}
\end{figure}

We begin by analysing the generation region and trade rates of a single player's optimal strategy in Figure~\ref{fig:boundary_single_base} {and Figure~\ref{fig:boundary_single_base_trade}, respectively}. When inventory is greater than the regulatory threshold $R = 5$, there is no generation region and the trading rate is negative as the player can sell off excess inventory for profit. When below the regulatory threshold and below the penalty value \$2.50, the player trades at a positive rate (i.e.~purchases OCs). {The generation region at the start of the period is below the penalty value and when inventory is less than the requirement. This region grows as time approaches $T$ and begins to include area above the penalty value. The corresponding trade rates in this region are negative, demonstrating that it is beneficial for the agent to generate and sell their inventory at a premium above the penalty rather than use them for regulatory purposes. We find positive trade rates for early time points when the player is above their requirement and the OC price is sufficiently low, as they anticipate the price will converge to the penalty and can sell their excess inventory for a profit.}

\begin{figure}[H]
\centering
\includegraphics[width=0.75\textwidth]{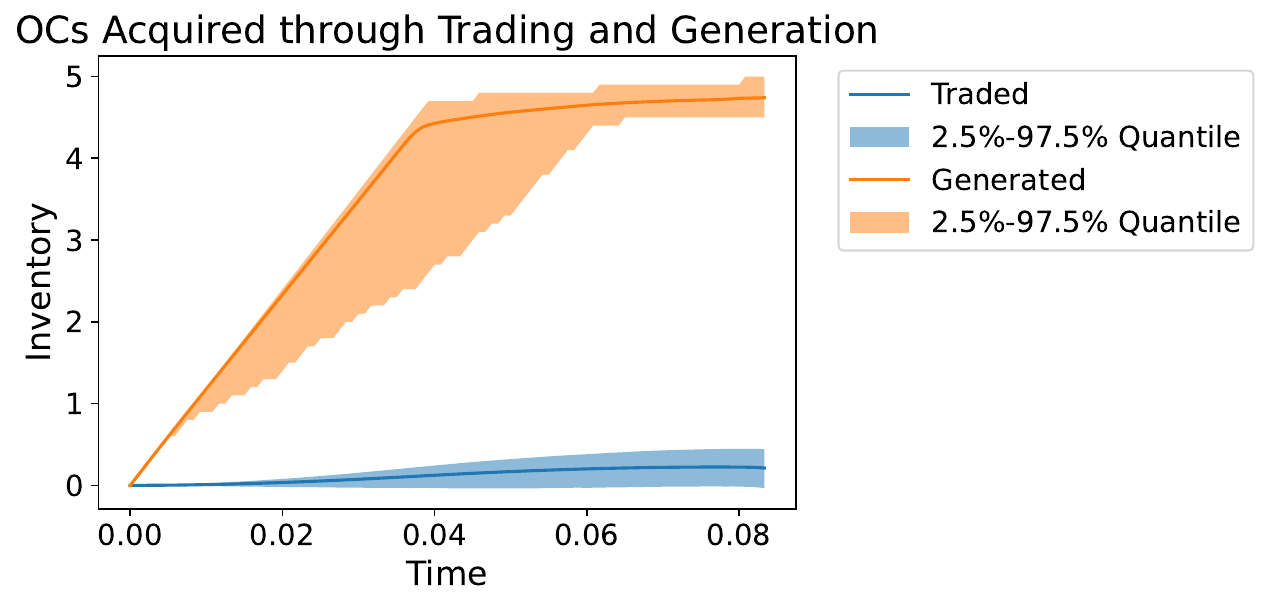}
\caption{Trade-Generation regions and trade rates at multiple time points. Generation region is in white.}
\label{fig:single_acquired_base}
\end{figure}

\begin{figure}[H]
\centering
\includegraphics[width=0.75\textwidth]{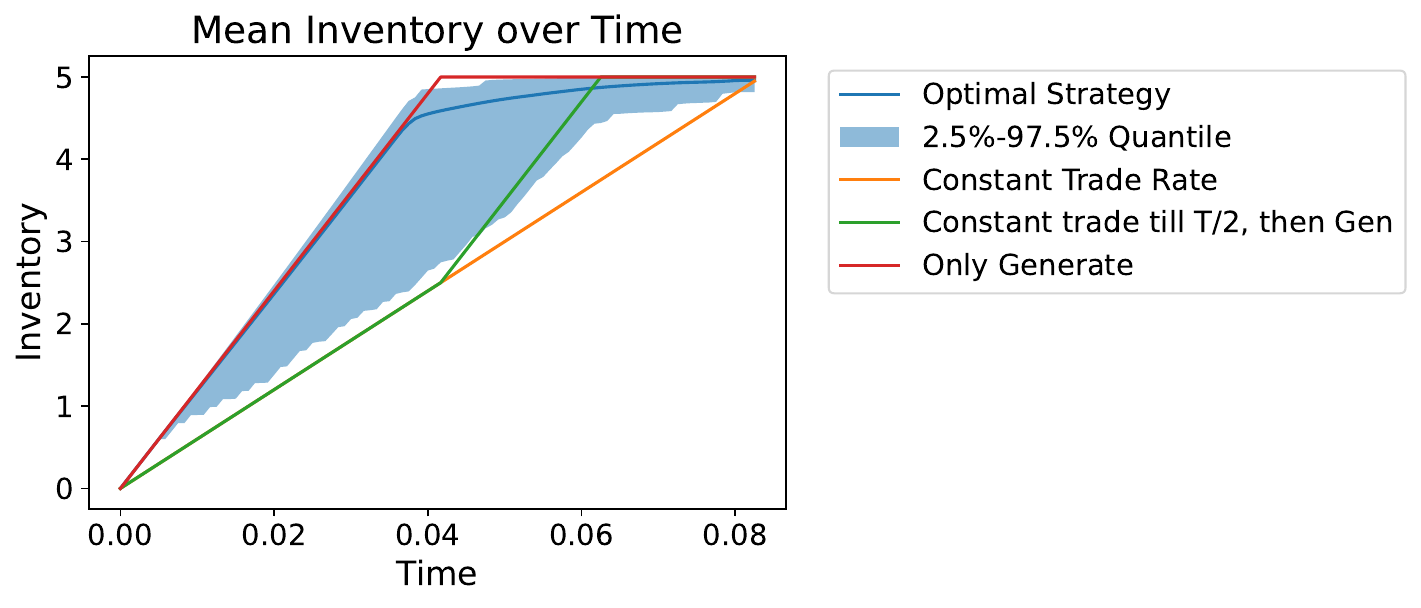}
\caption{Inventory with respect to time for the single-player optimal strategy and na\"{i}ve strategies with 95\% quantiles.}
\label{fig:naive_Inv}
\end{figure}

The player acquires the majority of their inventory through generation and only conducted minor trading, demonstrated in Figure~\ref{fig:single_acquired_base}. The player generates an average of {4.742} OCs throughout the period across all samples. As the initial spot price of an OC, $S_0$, is set to the penalty value of \$2.50, the player always begins the period in the generation region and stays in the generation region until price impact from successive generation creates a large enough impact to exit the generation region. When outside the generation region, small amounts of positive trading (buying) occurs. Figure~\ref{fig:naive_Inv} compares the mean inventories of the optimal strategy and three na\"{i}ve strategies. As the trade rates and generation instances in the na\"{i}ve strategies are fixed, no deviation occurs in their inventory over time.

\begin{figure}[H]
\centering
\qquad\includegraphics[width=0.95\textwidth]{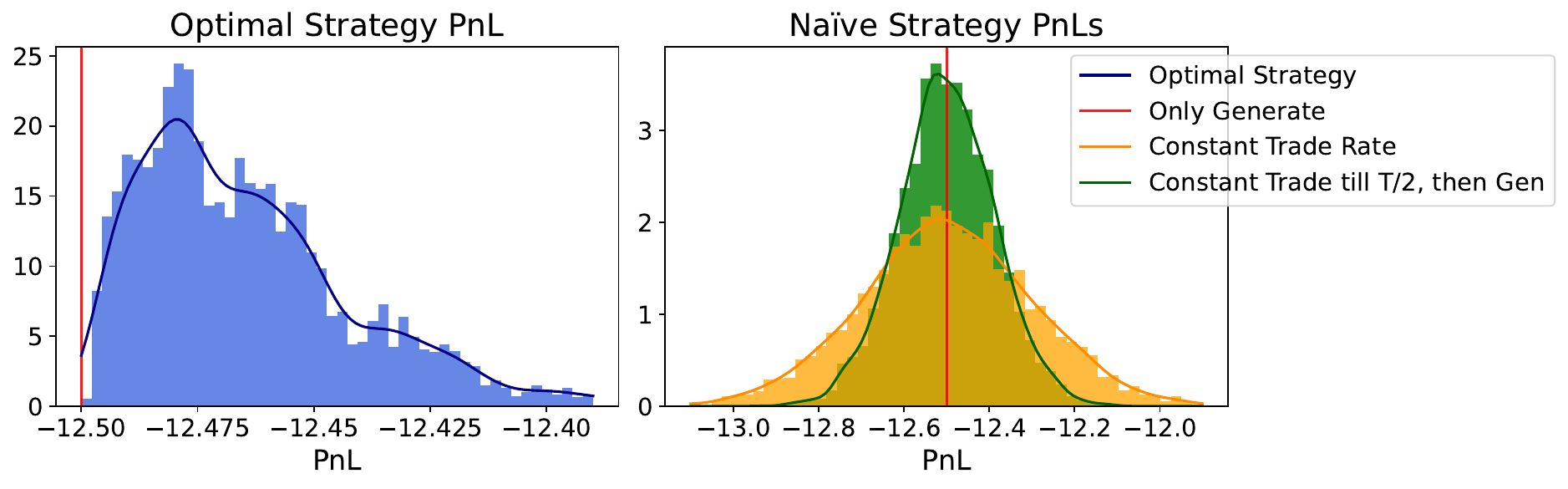}
\caption{Terminal PnL histograms for the single-player optimal strategy and na\"{i}ve strategies with kernel density estimates.}
\label{fig:naive_pnl}
\end{figure}

Each strategies' terminal profit--and--loss (PnL) is determined by combining the costs associated with each strategy and the terminal penalty. Figure~\ref{fig:naive_pnl} displays a histogram of each strategy's PnL along with its corresponding kernel density estimates. In all sample paths, the optimal strategy achieves a superior PnL to the only generate strategy, which is a constant value of $-\$12.50$. This is also the value of incurring the full penalty. While the na\"{i}ve strategies that include a trading component achieve better PnLs over the optimal strategy in a some sample paths, they achieve a worse expected PnL and {left tail expectation (TE)}. PnL and {TE} are presented in Table~\ref{tab:singleNaive_CVars}. When compared to all three na\"{i}ve strategies, the optimal strategy achieves both a superior mean PnL and {TE} demonstrating that the optimal strategy completely hedges against downside risk. {The optimal strategy's PnL standard error (SE), i.e.~the error in the estimator of the PnL's mean, is also much smaller than the other strategies that incorporate trading, indicating very little uncertainty in our ability to improve over the penalty.}

\begin{table}[H]
\centering
\begin{tabular}{ rrrrr } 
\toprule\toprule
 & Optimal  & Constant  & Half Trade, & Only  \\ 
 &Strategy &  Trade & Half Generate & Generate  \\
 \midrule
 Mean PnL & $-\$12.464$ & $-\$12.499$ & $-\$12.499$ & $-\$12.500$ \\ 
 {TE} & $-\$12.495$ & $-\$12.926$ & $-\$12.735$ & -- \\
 PnL SE ($\times10^{-3}$) & $0.349$ & $2.874$ & $1.582$ & -- \\
 \bottomrule\bottomrule
\end{tabular}
\caption{Mean PnL, {TE}, and PnL SE for the single-player optimal strategy and na\"{i}ve strategies.}
\label{tab:singleNaive_CVars}
\end{table}

\section{Two-Player Setting}\label{sec:twoPlayer}

In this section, we extend the single-player problem to the two-player game setting and characterise their equilibrium behaviour. With the inclusion of second player, the QVI from the single-player framework evolves into a bi-matrix game and we must apply a game theoretic approach (see~\cite{fudenberg1991game, osborne1994game, osborne2004introduction}) to analyse the resulting model. 

The Nash equilibrium, if it exists, may be a pure strategy or mixed strategy equilibrium. A pure strategy Nash equilibrium is one in which each, at each state, a player always chooses a specific action, and neither player can obtain a better value by deviating from their choice. A mixed strategy Nash equilibrium is one in which the players draw their strategy from a distribution over actions (independently of one another) -- i.e., their strategies are randomised. We will be seeking mixed strategy Nash equilibria. For a mixed strategy Nash equilibrium, a player's reward equals the sum over the action rewards multiplied by the probabilities that each player takes the appropriate actions.

In the bi-matrix game representing the GHG OC market, each player has {the choice whether or not to invest in OC generation}. We denote each players' value functions with a $1$ or $2$ to indicate ownership. Players may have differing terminal penalties, e.g., player one may require more credits to offset their excess emissions than player two. Therefore, we denote the terminal penalty of player $m$ as $G^{(m)}$, where $G^{(m)}(x) = - p\,(R^m - x)_+$. In the two-player setting, an additional complexity is introduced. Specifically, if one player trades while the other generates, the price impact affects both players. In this case, the trading player benefits when purchasing at a lower cost and is at a disadvantage when selling. If both players choose to generate OCs the price impact will be incurred twice. For this two-player setting we work directly in discrete time, as there is no hope for analytical tractability in the continuous time setting.

\subsection{Numerical Implementation}\label{sec:numImpTwo}

Extending the discrete time approach presented in Section~\ref{sec:numImpOne} to the two-player setting is straightforward. As in the single-player case, we denote the value function of the two possible player states, trading and generating as $U^{(m)\star}$ and $U^{(m)\dagger}$, $m\in\mfM:=\{1,2\}$. Using the same notation and framework as Section~\ref{sec:numImpOne}, for $t\in[t_{k-1},t_{k}]$, we define
\begin{equation}\label{eq:noGenCost_2p}
    U^{(m)\star}(t,\bx,s) = \sup_{\nu^{(m)}\in\mathcal{V}}\;\E_{t,\bx,s}\left[V^{(m)}(t_k,\bX_{t_k},S_{t_k}) - \int_t^{t_k} S_u\,\nu_u^{(m)}\, du - \frac{\kappa}{2}\int_t^{t_k} (\nu_u^{(m)})^2\, du \right]\;,
\end{equation}

where $\bx$ is the two-dimensional vector representing players' inventory. Note, the agents must track the inventory  of the other agent as well as their own inventory, hence the value functions depend on the vector of states.  
The PDE associated with~~\eqref{eq:noGenCost_2p} is the same as was found in the single-player framework, given by~\eqref{eq:oneP_U_HJB}, but with terminal conditions $U^{(m)\star}(t_k,\bx,s) = V^{(m)}(t_k,\bx,s)$ and the spatial derivative in $\partial_x$ replaced by $\partial_{x_m}$. Hence, similar to the single player setting, the optimal trading rate for player-$m$ is given in feedback form as
\begin{equation}
v^{*(m)}(t,\bx,s) = \frac{1}{\kappa}\left(\partial_{x_m} V^{(m)}(t,\bx,s) - s\right)\,.
\label{eqn:optimal-trading-rate-two-player}
\end{equation}

The value of each player at time $t = t_{k-1}$ is determined by the Nash equilibrium of the bi-matrix game shown in Table \ref{tab:game_sub}, where $\1_1:=(1,0)$ and  $\1_2:=(0,1)$. Players may not have the same inventory, therefore, we must compute the Nash equilibria for all possible combinations of the player's inventory.

\begin{table}[H]
\centering
\begin{tabular}{ cc|c|c| }
 & \multicolumn{3}{c}{\textbf{Player 2}}\\
  &  & Trade & Generate \\ 
 \cline{2-4}
 \multirow{3}{*}{\textbf{Player 1}} & Trade & \makecell{$U^{(1)\star}(t,\bx,s),$ \\ $~U^{(2)\star}(t,\bx,s)$} & \makecell{$U^{(1)\star}(t,\bx,s-\eta\,\xi_2),$\\$~{U^{(2)\star}}(t,\bx+\xi_2\1_2,s-\eta\,\xi_2) - \cxi_2$} 
 \\
 \cline{2-4}
 & Generate & \makecell{${U^{(1)\star}}(t,\bx+\xi_1\1_1,s-\eta \,\xi_1)- \cxi_1 ,$\\$~U^{(2)\star}(t,\bx,s-\eta\,\xi_1)$}   & \makecell{${U^{(1)\star}}(t,\bx+\xi_1\1_1,s-\eta \,(\xi_1+\xi_2))- \cxi_1,$\\$~{U^{(2)\star}}(t,\bx+\xi_2\1_2,s-\eta\,(\xi_1+\xi_2))- \cxi_2$} \\ 
 \cline{2-4}
\end{tabular}
\caption{\label{tab:game_sub}Bi-matrix game of two-player GHG OC market for a two-player setting.}
\end{table}

The Nash equilibrium of the game given by Table~\ref{tab:game_sub} can be determined by numerically estimating the solutions to four PDEs, as each player has two PDEs (continuation and holding) characterising their actions' values. To estimate the PDE solutions, we utilise implicit-explicit FD scheme developed in Section~\ref{sec:numImpOne}. The implicit FD matrix systems, given by~\eqref{eq:implicit_oneTrade}, has the same structure in the two-player setting, with the appropriate ownership subscript applied to the relevant terms. To compute each players' value from the resulting Nash equilibrium, we use the mixed strategy Nash equilibrium probabilities. We denote $\pi^{(m)}_{t,\bx,s}$ to be the probability associated with player--$m$ generating OCs at state $(t,\bx,s)$, which is found by solving for the Nash equilibrium in Table~\ref{tab:game_sub}. Letting $(-m)$ denote the other player in the game, the $m^{th}$ player's value is determined by
\begin{equation}\label{eq:value_nash}
    \begin{split}
    V^{(m)}(&t_{k-1},\bx,s) 
    \\
    =&\; (1-\pi^{(m)}_{t_{k-1},\bx,s})\,(1-\pi^{(-m)}_{t_{k-1},\bx,s})\,U^{(m)\star}(t_{k-1},\bx,s)
    \\&
    +(1-\pi^{(m)}_{t_{k-1},\bx,s})\,\pi^{(-m)}_{t_{k-1},\bx,s}\,U^{(m)\star}(t_{k-1},\bx,s-\eta\,\xi_{-m})
    \\
    &+\pi^{(m)}_{t_{k-1},\bx,s}\,(1-\pi^{(-m)}_{t_{k-1},\bx,s})\,{U^{(m)\star}}(t_{k-1},\bx+\xi_m\1_m,s-\eta\,\xi_m)
    \\
    &+\pi^{(m)}_{t_{k-1},\bx,s}\,\pi^{(-m)}_{t_{k-1},\bx,s}\,{U^{(m)\star}}(t_{k-1},\bx+\xi_m\1_m,s-\eta\,(\xi_m + \xi_{-m})).
    \end{split}
\end{equation}
The form of \eqref{eq:value_nash} is determined by the probabilities of each player choosing the action of trading or generating multiplied by the associated value corresponding to that action. 

The algorithm for solving for the mixed-strategy Nash equilibria in this environment is given in Algorithm \ref{algo:two-player}. We utilise the  Python library \texttt{NashPy} to solve for the Nash equilibria. To improve computational efficiency, one can use parallel programming to simultaneously compute the equilibria.

\begin{algorithm}[H]
	\caption{Two-player value functions and mixed strategy Nash equilibria}\label{algo:two-player}
    \footnotesize
	\KwIn{$t,s,\bx$ grid}
    Set terminal values $V^{(m)}_{N,\bi,j}=G^{(m)}(x_{\bi_m})$ for $m\in\mfM$\;
    \For{$k = N, N-2, \ldots,1$}{
    \For{each $i_1,i_2\in\mfI$}{
        Solve \eqref{eq:implicit_oneTrade} for $U^{(m)\star}_{k-1,\bi,\cdot}$ for each  $m\in\mfM$ using interpolation for boundary points in $\mfI$\;
        Compute the mixed-strategy Nash equilibria for the bi-matrix game in Table \ref{tab:game_sub} to calculate and store $V^{(m)}_{{k-1},i,j}$ and $\pi^{(m)}_{{k-1},\bi,j}$ for all $j\in\mfJ$, $m\in\mfM$\;
        Store the optimal trading rate $v^{*(m)}_{k-1,\bi,j}$ using the FD approximation of \eqref{eqn:optimal-trading-rate-two-player} for all $j\in\mfJ$, $m\in\mfM$\;    
    }

    }
    \KwOut{$V^{(m)}$, $v^{*(m)}$, and $\pi^{(m)}$ for $m\in\mfM$}
\end{algorithm}

\subsection{Results}\label{sec:resTwo}

\subsubsection{Homogeneous Investment Opportunities}\label{sec:homo_opps}

We first analyse two regulated players that have equal investment opportunity for OC generation ($\xi_1 = \xi_2$ and $\cxi_1 = \cxi_2$). The values of $\xi$ and $\cxi$ are chosen such that the cost of generating one OC is equal to the penalty value. We use a linearly spaced grid for the spatial coordinates over $100$ time points, with simulation parameters presented in Table~\ref{tab:base_params_two} and we simulate $5,000$ OC price sample paths to examine a player's optimal strategies. Each player begins the period with zero inventory (i.e~$X_{0,\,i} = 0$).

\begin{table}[H]
\centering
\begin{tabular}{ ccccccccc } 
\toprule\toprule
 T (years)  & $\sigma$ & $\kappa$ & $\eta$ & $\xi_{1,2}$ & $\cxi_{1,2}$ & $S_0$ & $R^{1,2}$ & Penalty \\ 
 \midrule
 1/12 & 0.5 & 0.03 & 0.05 & 0.1 & 0.25 & 2.5 & 5 & 2.5 \\ 
 \bottomrule\bottomrule
\end{tabular}
\caption{Simulation parameters for a homogeneous two player market.}
\label{tab:base_params_two}
\end{table}

\begin{figure}[h!tbp]
    \centering
    \begin{minipage}{0.75\textwidth}
        \centering
        \includegraphics[width=\textwidth]{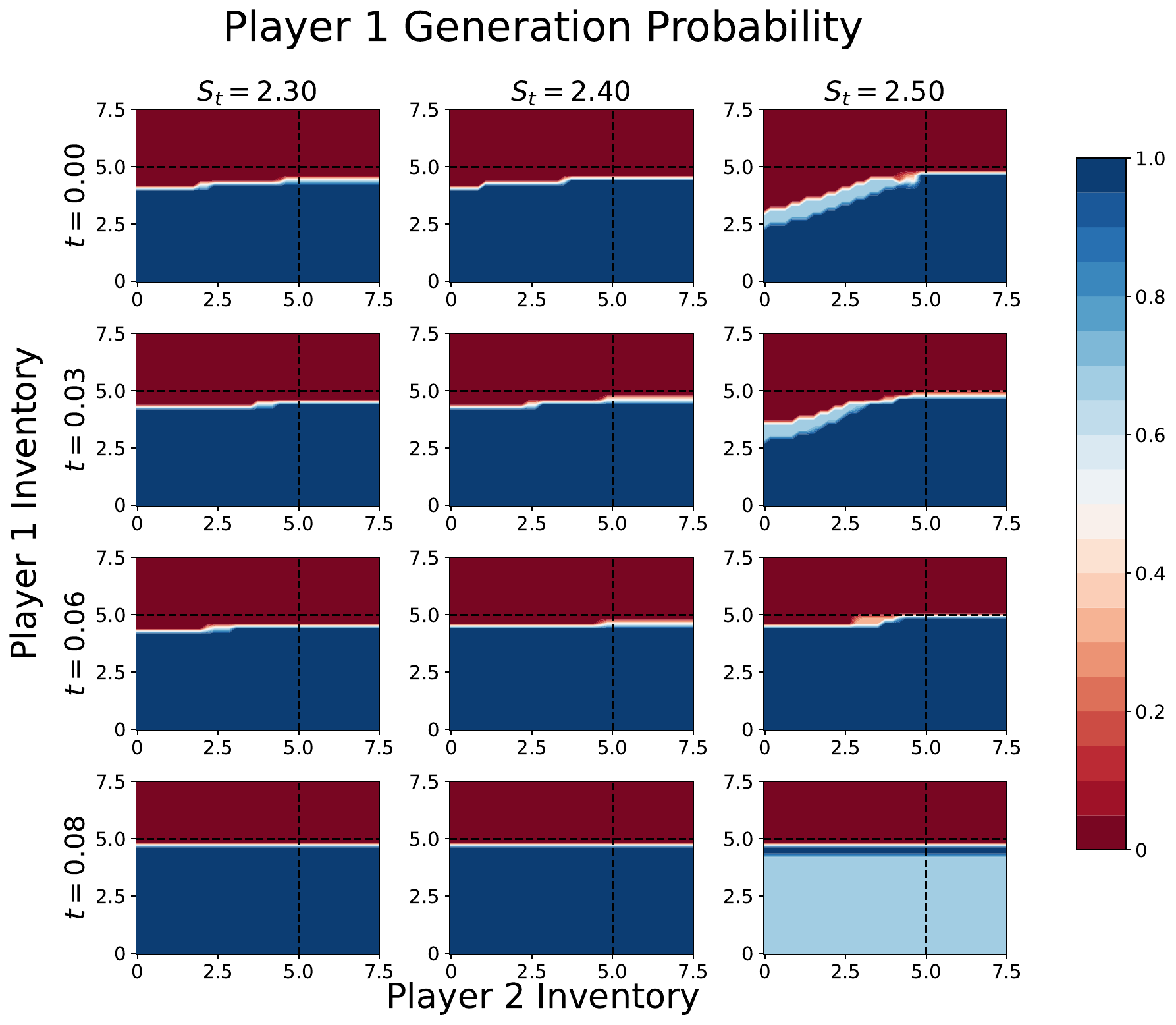} 
    \end{minipage}\vspace{0.25cm}
    \begin{minipage}{0.75\textwidth}
        \centering
        \includegraphics[width=\textwidth]{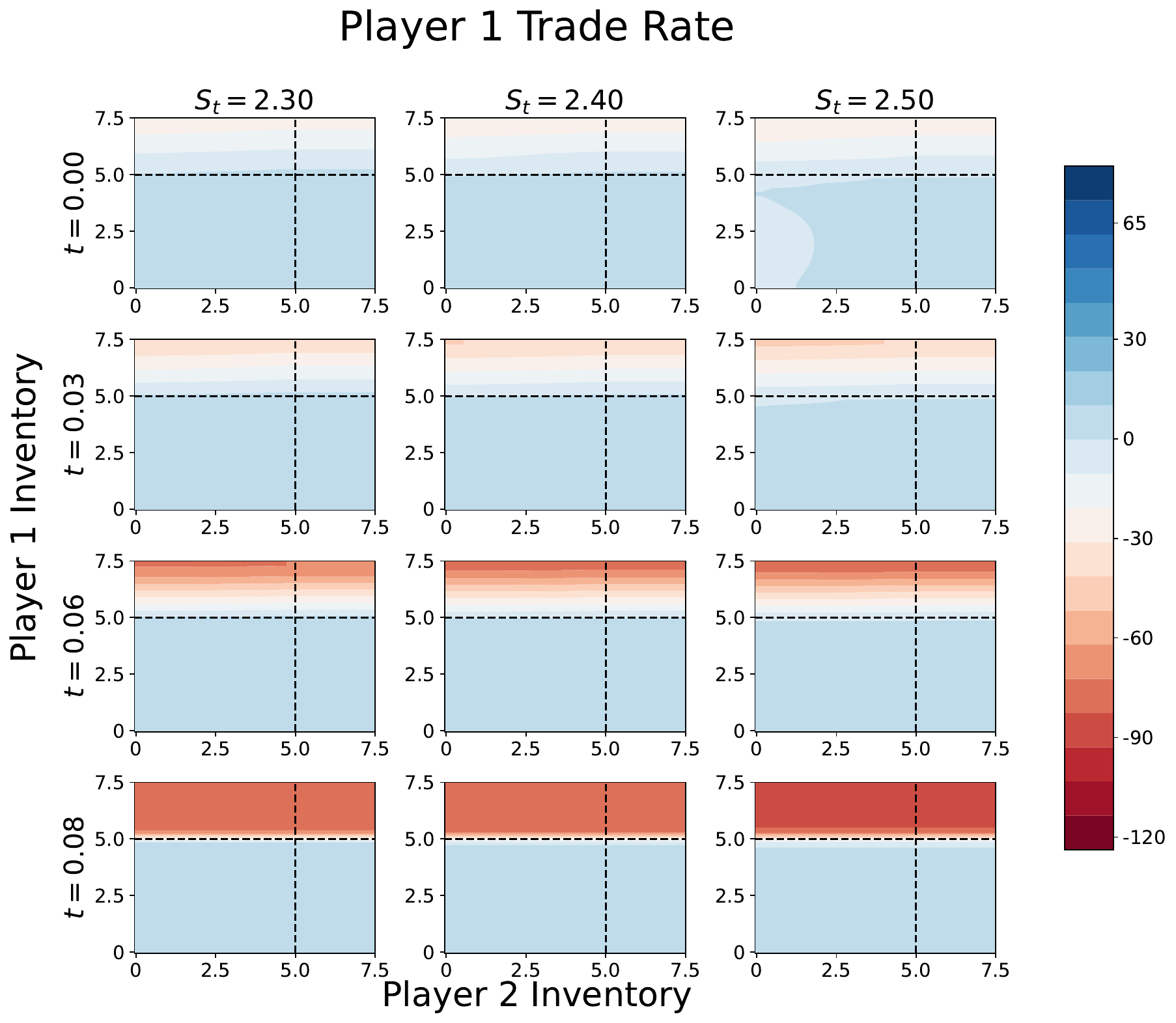} 
    \end{minipage}
    \caption{Trading probability and trade rate conditional on taking a trading action of player one. Rows represent time points throughout the period and columns are OC prices. Player two's results are equivalent to player one when rotated 90 degrees.}
    \label{fig:nash_probs}
\end{figure}

Figure~\ref{fig:nash_probs} displays the probabilities that player one will generate at a given state in the upper panel and their trade rates in the lower panel. Due to homogeneity, player two has identical probabilities and trade rates. 
Areas of high generation probability (low trading probability) occur earlier in the period and when the player's inventory is below their requirement {and the OC price is below the penalty value. As time approaches $T$, the boundary between low and high generation probability becomes linear along the respective player's terminal inventory requirement. The optimal trade rates are intuitive:  a player purchases OCs when their inventory is below the requirement, and as time approaches $T$  they sell excess inventory at faster rates. We find that at early time points,  players do purchase OCs when their inventory exceeds the requirement and the price is below the penalty value, as players anticipate the OC price will increase to the penalty by $T$. If players start the period with zero OCs, they begin in a region of generation and subsequently impact the OC price pushing it down. As such, our figures focus on OC prices that are at or below \$2.50, as that is the region the players  the environment remains when following the optimal strategy.} 

\begin{figure}[H]
\centering
\begin{minipage}{.8\textwidth}
  \centering
  \includegraphics[width=\textwidth]{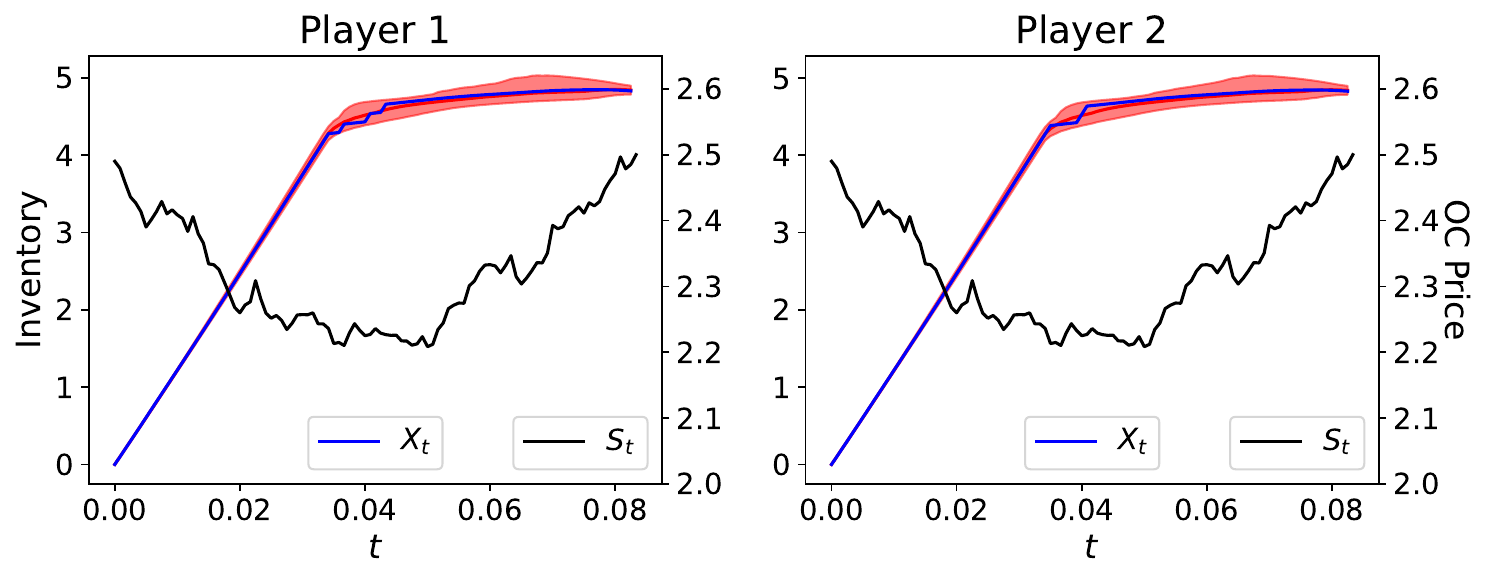}
\end{minipage}\hfill
\begin{minipage}{.8\textwidth}
  \centering
  \includegraphics[width=\textwidth]{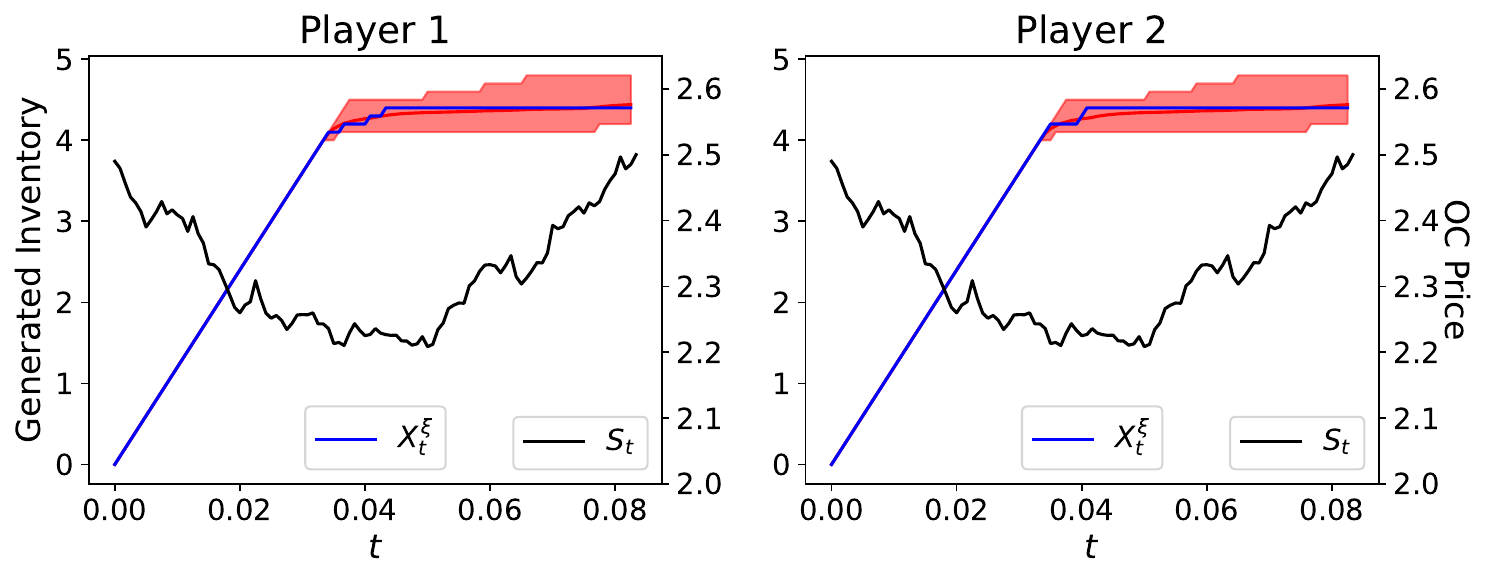}
\end{minipage}
\caption{Total inventory (top) and generated inventory (bottom) with mean inventory (red), 95\% quantiles (shaded red), and sample path (blue) for a homogeneous players. Corresponding OC price sample path (black) on right axis.}
\label{fig:homog_inv}
\end{figure}

Figure~\ref{fig:homog_inv} displays  how  players' inventories evolve over the period and the inventory they acquire through generation. {As in the single player scenario, both players generate the majority of their OCs in a period of successive generation at the start of the period, driving the OC price down due to market impact. This is followed by a period of trading with sporadic generation instances when the players are able to trade at a more advantageous OC price. In the figure, the sample paths and quantiles are extremely similar due to  symmetric nature of the players --  only minor differences occurring due to small differences when each player generates at the midway point due to their probabilities.}

The terminal PnL distribution of both players, shown in Figure~\ref{fig:PnL2_equal}, demonstrates that both players achieved non-trivial improvement over the maximum terminal penalty of $-\$12.50$ and the PnLs of the na\"{i}ve strategies from Section~\ref{sec:resOne}. Each players' mean PnL and {TE} are given in Table~\ref{tab:equalMeans}. As the players are homogeneous, they obtain the same summary statistics. Both players completely hedge against downside risk when behaving in accordance to the Nash equilibria, as demonstrated by their {TE}.

\begin{minipage}{0.45\textwidth}
\centering
\footnotesize
\begin{tabular}{ cccc } 
\toprule\toprule
  & & & PnL SE 
  \\
  & Mean PnL & {TE} & ($\times10^{-3}$)  \\ 
 \midrule
 Player 1 & $-\$12.393$ & $-\$12.467$ & 0.622 \\
 Player 2 & $-\$12.393$ & $-\$12.467$ &  0.622 \\ 
 \bottomrule\bottomrule
\end{tabular}
\captionof{table}{Mean PnL, {TE}, and PnL SE for homogeneous players.}
\label{tab:equalMeans}
\end{minipage}
\hfill
\begin{minipage}{0.45\textwidth}
\centering
\includegraphics[width=0.9\textwidth]{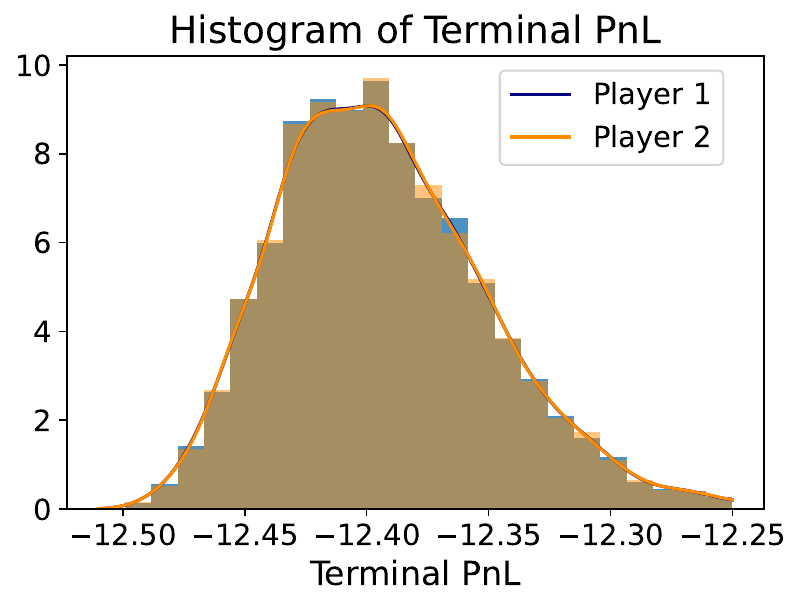}
\captionof{figure}{Terminal PnL histograms and KDEs of players with homogeneous investment opportunities.}
\label{fig:PnL2_equal}
\end{minipage}

\subsubsection{Heterogeneous Investment Opportunities}\label{sec:hetero}

Next, we explore players who have heterogeneous investment opportunities (i.e.~generation capacity), such that~$\xi_1 \neq \xi_2$ and $\cxi_1 \neq \cxi_2$. This setting represents firms that have different project investment opportunities that generate an unequal amount of instantaneous OCs, and can representing a small firm and a large firm. The remaining market parameters remain as in Table~\ref{tab:base_params_two}, with the modifications: $\xi_1 = 0.1,~\xi_2 = 0.4,~\cxi_1 = \$0.25,~\text{and}~\cxi_2 = \$1.00$. As before, the marginal price of generation is equal to the penalty value for both players and we generate $5,000$ OC price samples paths to analyse their optimal strategies. In all paths, both players start with an initial inventory of zero OCs.

\begin{figure}[h!tbp]
    \centering
    \begin{minipage}{0.8\textwidth}
        \centering
        \includegraphics[width=\textwidth]{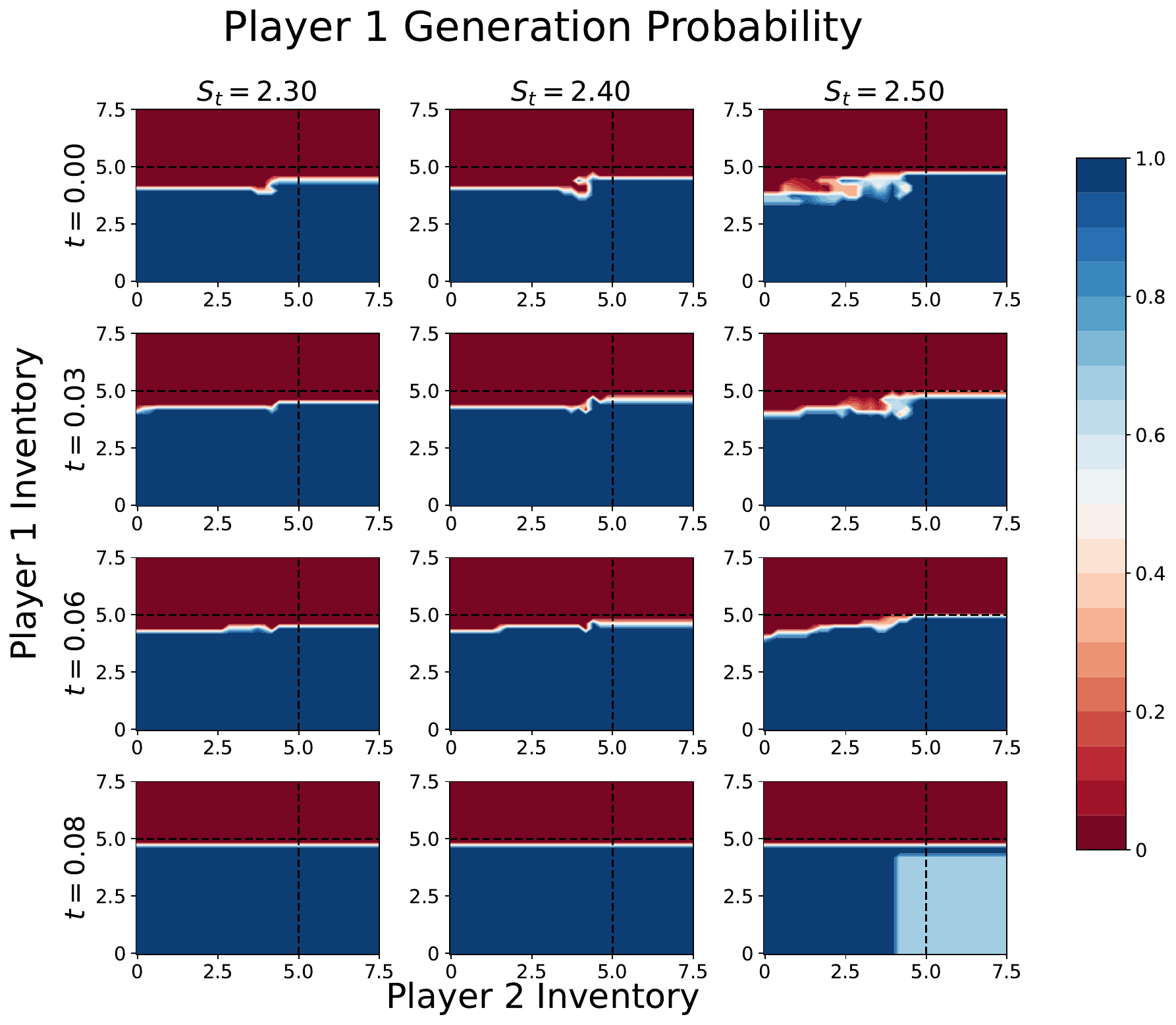} 
    \end{minipage}\vspace{0.25cm}
    \begin{minipage}{0.8\textwidth}
        \centering
        \includegraphics[width=\textwidth]{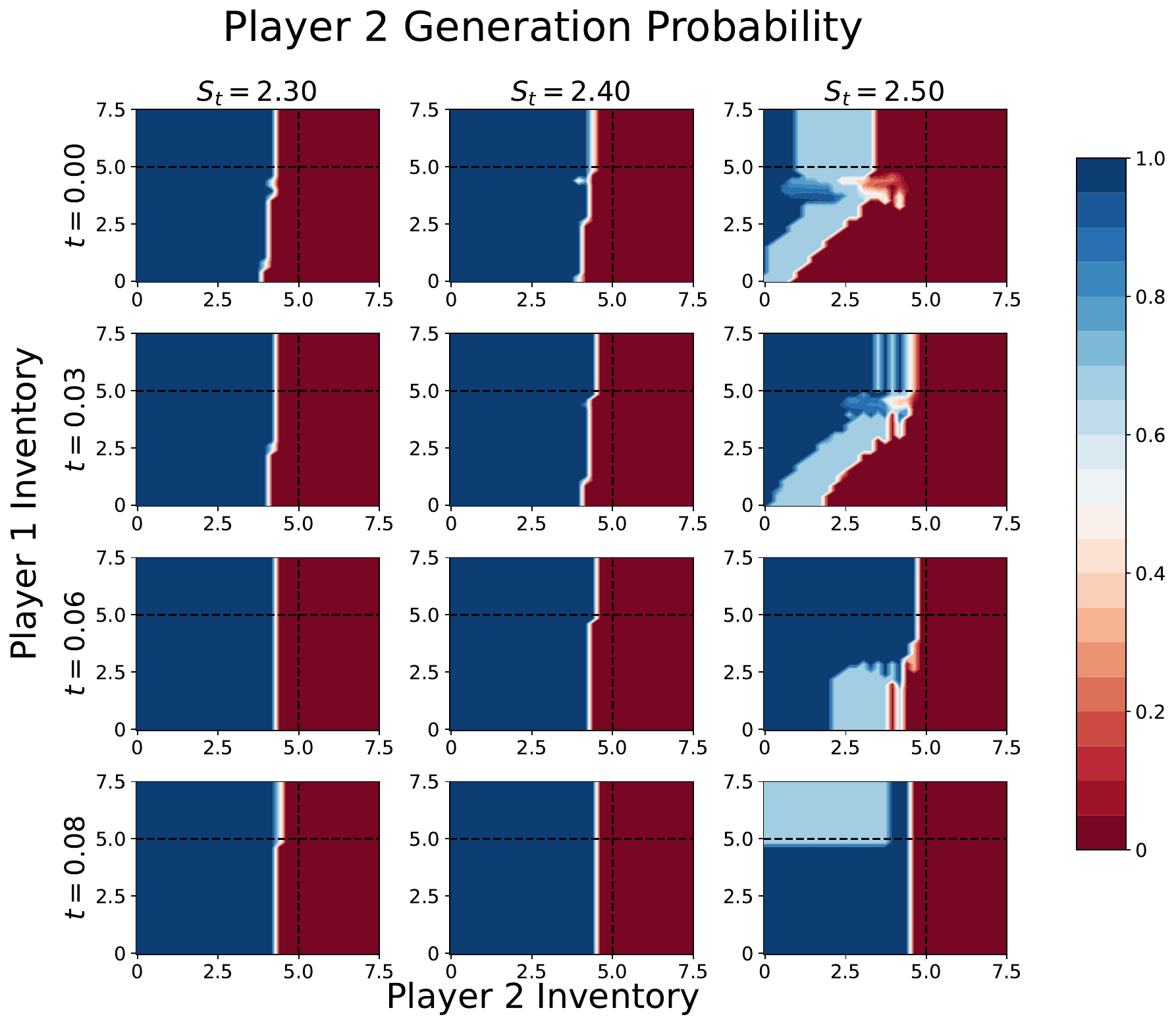} 
    \end{minipage}
    \caption{Trading probability of players with heterogeneous generation capability. Rows represent time points throughout the period and columns are OC prices.}
    \label{fig:hetero_probs}
\end{figure}

The players' Nash equilirbia probabilities for generating are presented in Figure~\ref{fig:hetero_probs}.
\begin{figure}[h!tbp]
    \centering
    \begin{minipage}{0.75\textwidth}
        \centering
        \includegraphics[width=\textwidth]{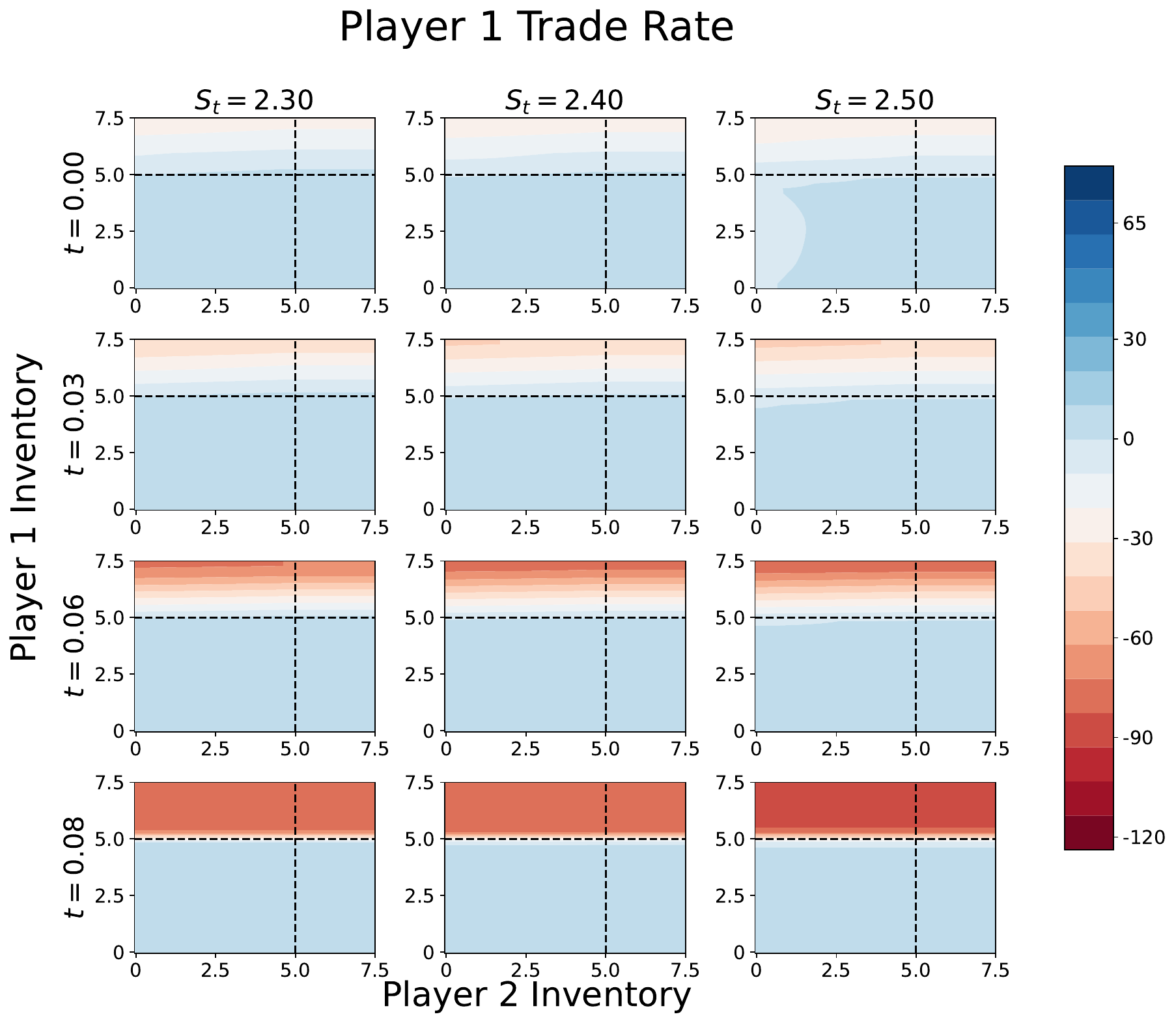} 
    \end{minipage}\vspace{0.25cm}
    \begin{minipage}{0.75\textwidth}
        \centering
        \includegraphics[width=\textwidth]{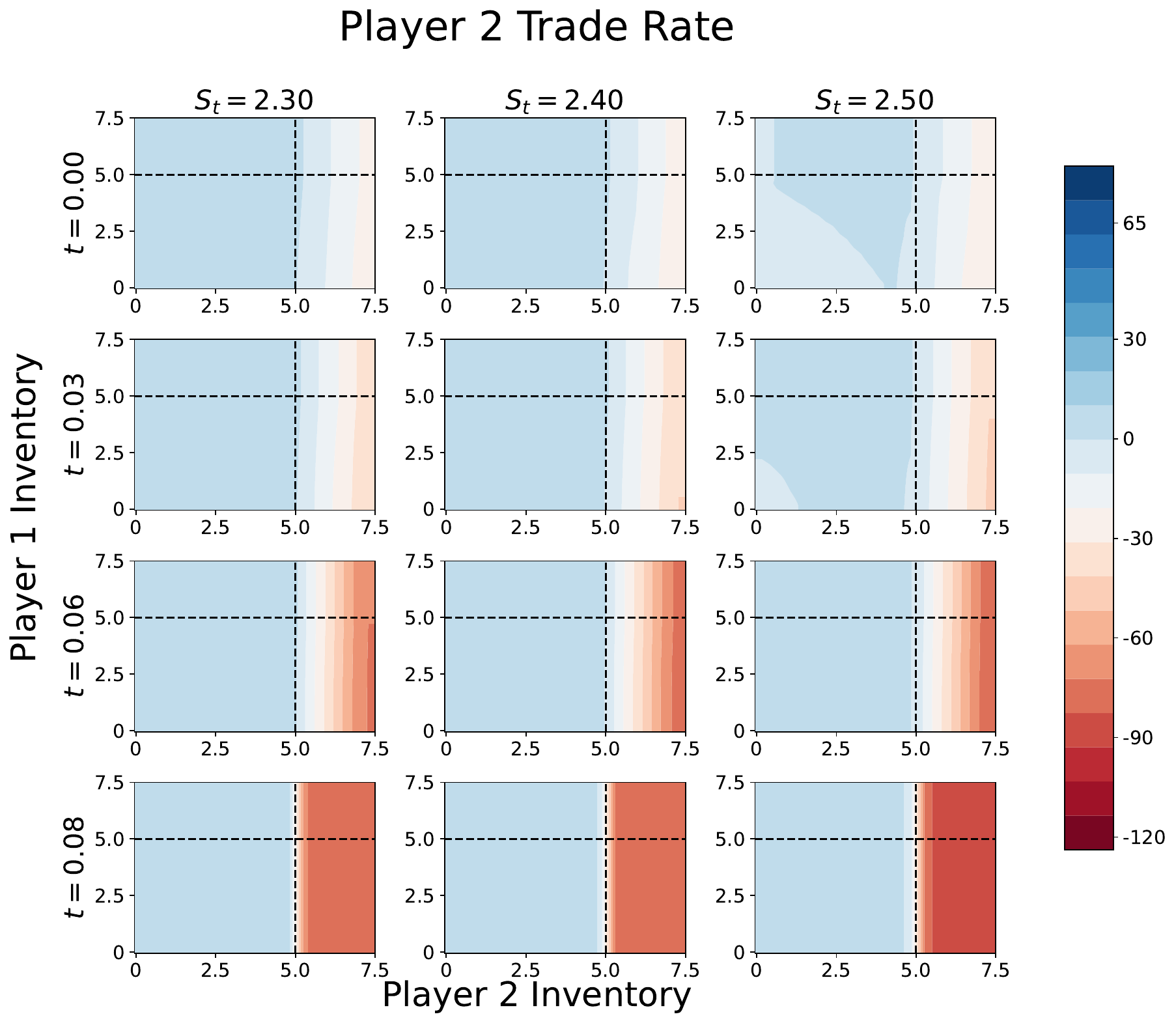} 
    \end{minipage}
    \caption{Trade rates conditional on taking a trading action of players with heterogeneous generation capability. Rows represent time points throughout the period and columns are OC prices.}
    \label{fig:two_trade_unequal}
\end{figure}
Player one's generation probabilities are very similar to those of the homogeneous case presented in the previous section. Player one has large regions of high generation probability when below their OC threshold and when at or below the penalty value. Player two has more variability in their generation probability than player one, due to their advantage in their generation capacity. As they can generate four times the amount of OCs than player one at any time their strategy allows for more freedom in their action choice, represented by the probabilities that are not concentrated at zero or one. {For both players, we find there is more variability in the generation probabilities when the OC price is at the penalty value (which equals the marginal cost to generate one OC).}

The corresponding trade rates are displayed in Figure~\ref{fig:two_trade_unequal}. These results are similar to the homogeneous player setting. Trade rates are negative when player's inventories exceed the requirement and positive when below the requirement. For OC prices that exceed the penalty, minor amounts of negative trading take place as it is more advantageous for players to sell their inventory at the premium over the penalty value and incur the penalty.

\begin{figure}[H]
\centering
\includegraphics[width=0.85\textwidth]{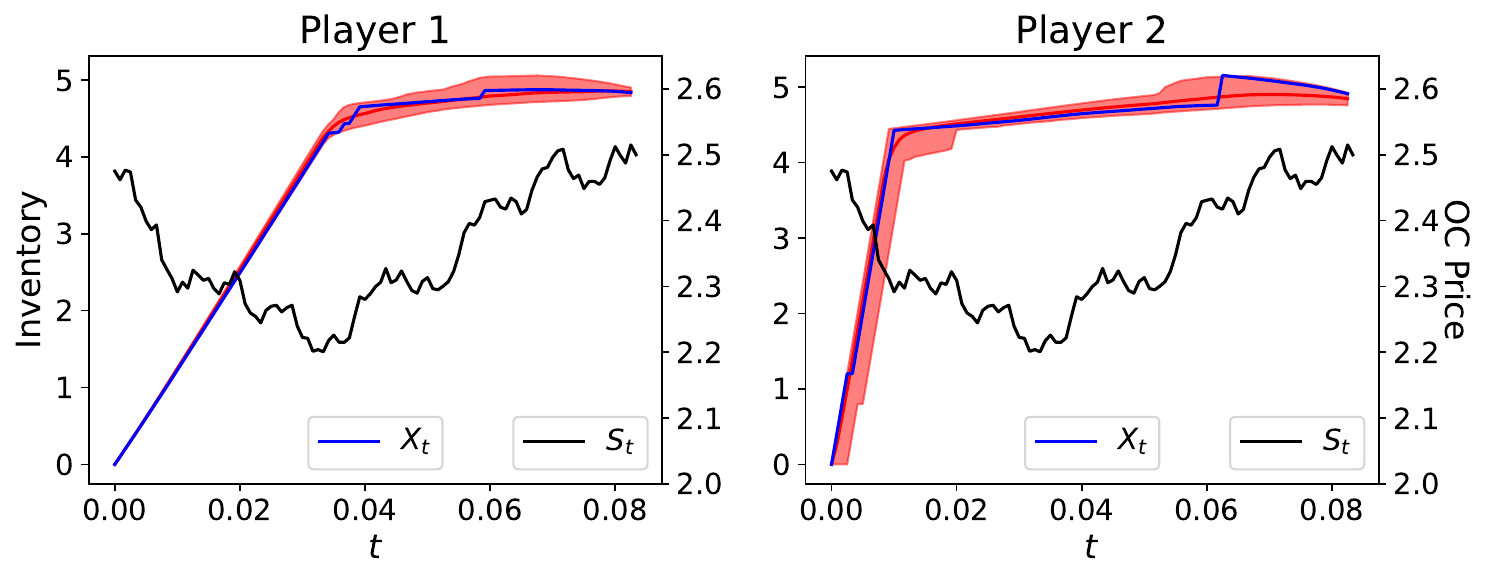}
\caption{Mean inventory (red), 95\% quantile (shaded red), and sample path (blue) on left axis and corresponding OC price sample (black) on right axis for heterogeneous players.}
\label{fig:inventory_larger}
\end{figure}

\begin{figure}[H]
\centering
\includegraphics[width=0.85\textwidth]{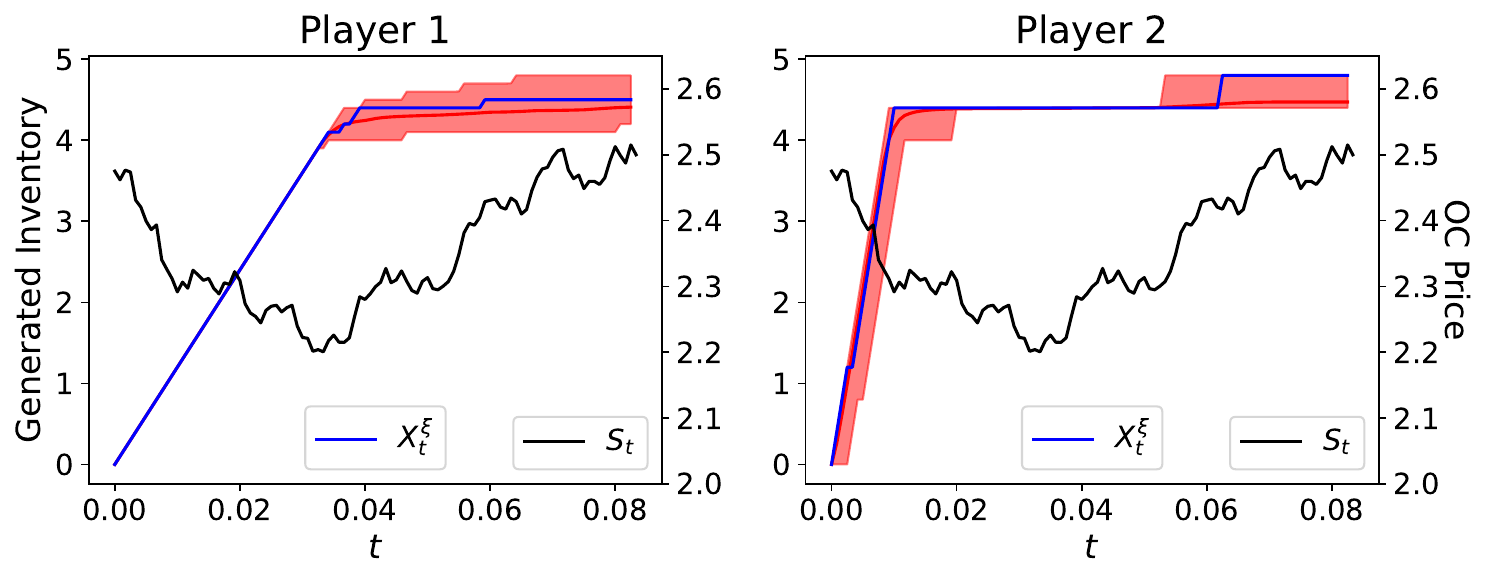}
\caption{Mean inventory acquired by generation (red), 95\% quantile (shaded red), and sample path (blue) on left axis and corresponding OC price sample (black) on right axis for heterogeneous players.}
\label{fig:inventory_gen_larger}
\end{figure}

Figures~\ref{fig:inventory_larger} and~\ref{fig:inventory_gen_larger} demonstrate the difference in the strategies that players take. Both players begin with successive instances of generation. Player two then begins a long period of trading with intermittent generation. Meanwhile, player one has a period of trading {beginning around $\frac12 T$  with only minor generation instances. The successive generation by both players drives down the OC price, demonstrated by the sample path (black), creating a mutually beneficial environment where both players can trade at more advantageous prices. Both players generate the majority of their required OCs.}

Figure~\ref{fig:PnL2_larger} displays the terminal PnL histograms of each player while Table~\ref{tab:heteroMeans} displays the mean PnLs and CVaRs$_{95}$. For both players, the terminal PnL is always better than incurring the full penalty and their downside risk is completely eliminated. Both players achieve significantly better terminal statistics than the na\"{i}ve strategies from Section~\ref{sec:resOne}. {Player one achieves slightly better terminal statistics than player two, which we attribute to their lower costs to generate OCs. There is a modest increase in player one's SE of their PnL than there is with player two. Both players achieve superior terminal statistics than their homogeneous counterparts however, demonstrating the mutual benefits of interacting with  players with larger generation capacity.}
\\[1em]
\begin{minipage}{0.45\textwidth}
\centering
\hfill
\begin{tabular}{ cccc } 
\toprule\toprule
   & & & PnL SE \\
   &  Mean PnL & {TE} & ($\times10^{-3}$)  \\
 \midrule
 Player 1 & $-\$12.376$ &  $-\$12.456$ &  0.665 \\
 Player 2 & $-\$12.387$ & $-\$12.461$  & 0.545 \\ 
 \bottomrule\bottomrule
\end{tabular}
\captionof{table}{Mean PnL, {TE}, and PnL SE for heterogeneous players.}
\label{tab:heteroMeans}
\end{minipage}
\hfill
\begin{minipage}{0.45\textwidth}
\centering
\includegraphics[width=0.9\textwidth]{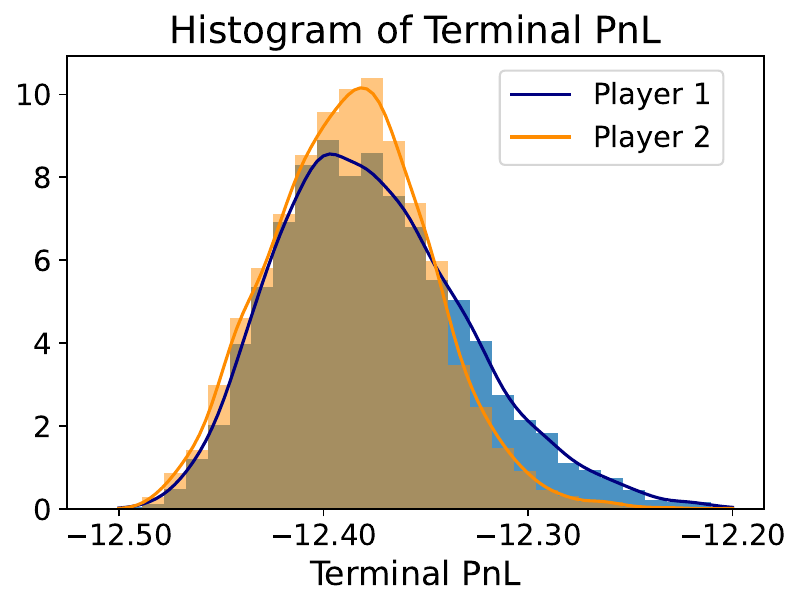}
\captionof{figure}{Terminal PnL histograms of players with heterogeneous investment opportunities.}
\label{fig:PnL2_larger}
\end{minipage}

\subsection{Comparisons}

In all prior simulations, model parameters and players' marginal cost of OC generation were equivalent. This allows for direct comparison across scenarios. Player's abiding by the optimal strategy always achieve a superior PnL compared to the full penalty and completely eliminate their downside risk. In other words, the {TE} values of all players are superior to the full penalty value. Given our parameter choices, all experiments demonstrate the benefit and importance of being able to generate OCs through project investment, as the majority of OCs are acquired by generation. Given the low liquidity of OCs in the Canadian market (\cite{sadikman2022evolution}) this result is in line with how firms currently behave. In Table~\ref{tab:gen_OC_Means}, the mean value of generated OCs across the simulations for each experiment is shown, alongside mean PnL and {TE}. Players across all experiments on average generate at least $85\%$ of their terminal OC inventory. Players that participate in the two-player market achieve a superior mean PnL and {TE} than the player who participates in the single-player market, demonstrating importance to model this style of market as a multi-player game. {Both players in the heterogeneous scenario achieve superior PnLs and TEs compared to their homogeneous counterpart, demonstrating an inherited benefit from participating in a market with players with larger generation capabilities. Across both two player settings, the players obtain similar statistics as they are allowed to simultaneously trade and generate OCs, which partially offsets some of the benefits of having a larger generation capacity.}

\begin{table}[H]
\centering
\begin{tabular}{ c||c||cc||cc } 
\toprule\toprule
 Player (Experiment) & Single Player & P1 (Equal) & P2 (Equal) & P1 (Hetero) & P2 (Hetero)  \\ 
 \midrule
 Generation Capacity & $\xi = 0.1$ & $\xi_1 = 0.1$ & $\xi_2 = 0.1$ & $\xi_1 = 0.1$ &$\xi_2 = 0.4$  \\ 
 \midrule
 Mean Generated OCs & 4.742 & 4.444 & 4.443 & 4.412 & 4.473 \\ 
 Mean PnL & $-\$12.464$ & $-\$12.393$ & $-\$12.393$ & $-\$12.376$ & $-\$12.387$ \\ 
 \
{TE} & $-\$12.495$ & $-\$12.467$ & $-\$12.467$ & $-\$12.456$ & $-\$12.461$ \\ 
 \bottomrule\bottomrule
\end{tabular}
\caption{Mean generated OCs, PnL, and {TE} for each single-period experiment.}
\label{tab:gen_OC_Means}
\end{table}

\subsection{Multi-Period Model}\label{sec:multi_period}

OCs in the Canadian market last for multiple years (and compliance periods) if they are not submitted for compliance purposes. Hence, it is natural to extend our methodology to accommodate a multi-period framework. We assume a model with $L$--many compliance dates, where a compliance date takes place on $(T_l)_{l\in\{1,\dots,L\}}$ where the $l^{th}$ compliance period takes place between $[T_{l-1}, T_l)$ for $l\in\{1,\dots,L\}$ and $T_0$ represents the start of the first period. OCs acquired prior to compliance date $T_l$ may be used for regulatory purposes on $T_l$. Player-$m$'s assessment of a strategy $(\nu,\btau)$ in this setting (at state $(t, \bx, s)$) is given by
\begin{multline}
    J^{(m)\nu,\btau}(t,\bx,s)
    \\
    \;\;:=\E_{t,\bx,s}\!\left[
    \sum_{l=1}^L \ind_{t\leq T_l}\, G^{(m)}_l(X^{(m)}_{T_l}) - \int_t^{T_L} \big(S_u\,\nu^{(m)}_u  - \tfrac{\kappa}{2} (\nu^{(m)}_u)^2\, \big)du - \sum_{i\in\N} \ind_{t\le\tau_i\le T_L}\,\cxi_{m}
    \right]\!,
\end{multline}
where $G_l^{(m)}(x) = -p\,(R^{(m)}_l-x)_+$ such that $R^{(m)}_l$ is player--$m$'s OC requirement for the $l^{th}$ period to completely eliminate their penalty. As in the two-player setting, players' optimal values are determined by the Nash equilibrium induced by the bi-matrix game. The bi-matrix game structure and numerical implementation remains the as in the single-period two-player model from Section~\ref{sec:numImpTwo}, with minor modifications to allow for multiple compliance dates. 

The simulation parameters, presented in Table~\ref{tab:multi_period_params}, are similar to those in Section~\ref{sec:hetero} for two heterogeneous players. We analyse a scenario with two compliance periods with $T_1= \frac1{12}$ and $T_2=\frac2{12}$, representing a compliance date at the end of each month for two consecutive months. The trading friction is also increased, symbolising an environment that is more hostile to trading. We assume that players have the same OC requirement and this requirement is constant across both periods. We run the experiment using 150 time points, such that there are 75 time points for each period.

\begin{table}[H]
\centering
\begin{tabular}{ ccccccccccc } 
\toprule\toprule
 T (years)  & $\sigma$ & $\kappa$ & $\eta$ & $\xi_{1}$ & $\xi_{2}$ & $\cxi_1$ & $\cxi_2$ & $S_0$ & $R^{1,2}_{1,2}$ & Penalty \\ 
 \midrule
 $[1/12,\,2/12]$ & 0.5 & 0.06 & 0.05 & 0.1 & 0.4 & 0.25 & 1.00 & 2.5 & 5 & 2.5 \\ 
 \bottomrule\bottomrule
\end{tabular}
\caption{Simulation parameters for two-period two-player market with heterogeneous investment opportunities.}
\label{tab:multi_period_params}
\end{table}

\begin{figure}
    \centering
    \begin{minipage}{0.75\textwidth}
        \centering
        \includegraphics[width=\textwidth]{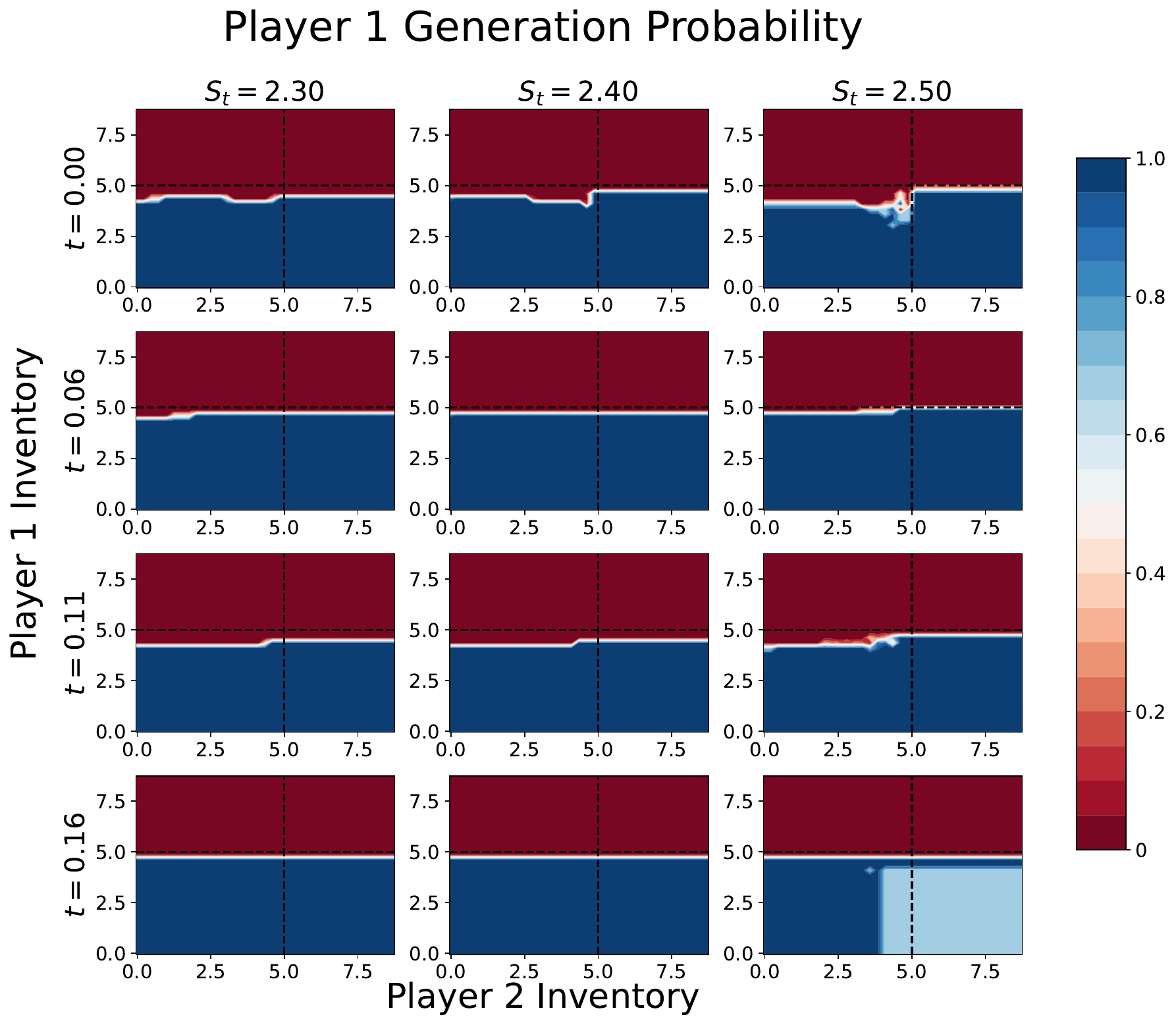} 
    \end{minipage}\vspace{0.25cm}
    \begin{minipage}{0.75\textwidth}
        \centering
        \includegraphics[width=\textwidth]{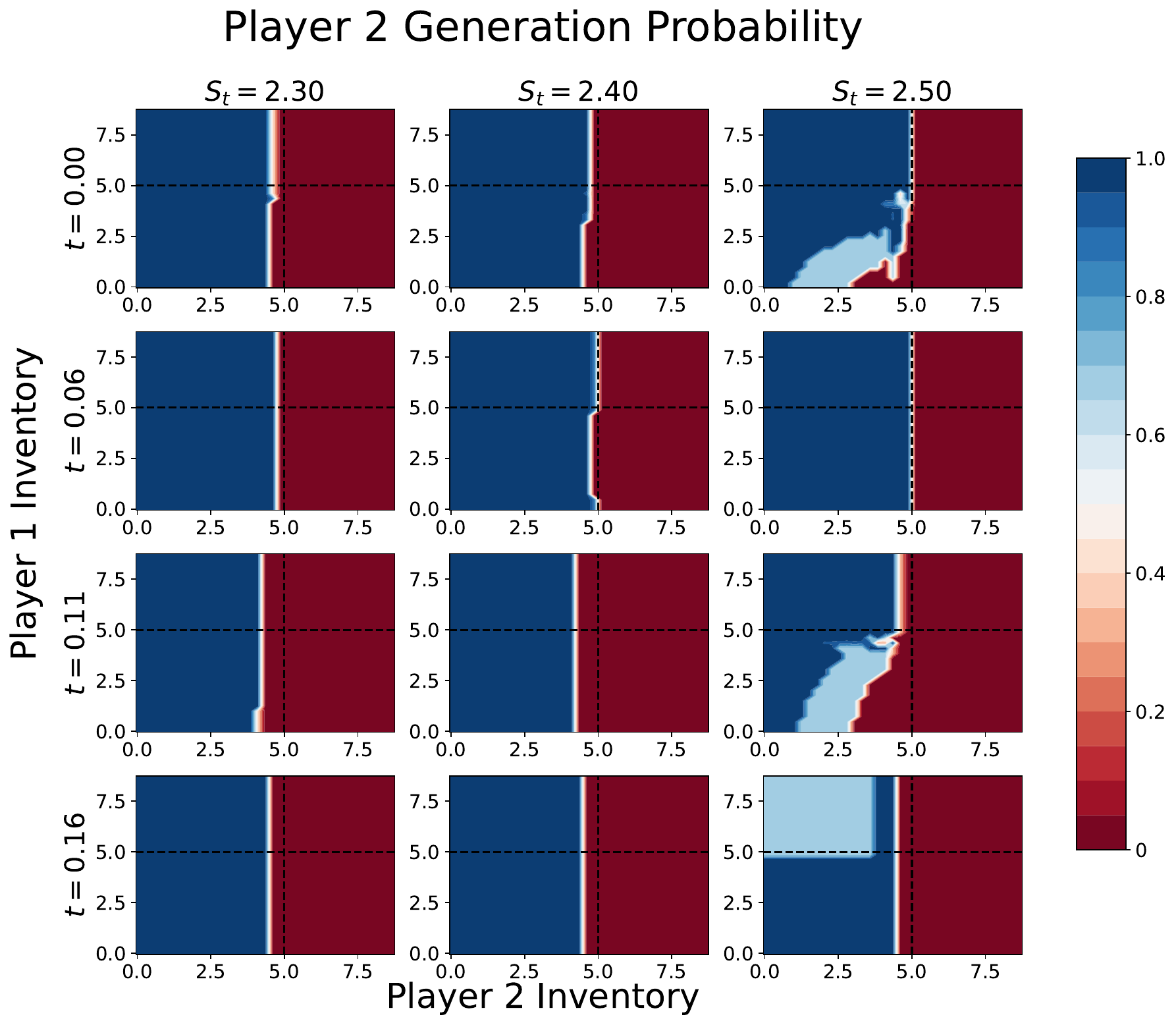} 
    \end{minipage}
    \caption{Trading probability of players with heterogeneous generation capability in a two-period setting. The first two rows in each sub-figure correspond to the first period while the last two rows take place in the second period.}
    \label{fig:multi_probs}
\end{figure}

\begin{figure}
    \centering
    \begin{minipage}{0.75\textwidth}
        \centering
        \includegraphics[width=\textwidth]{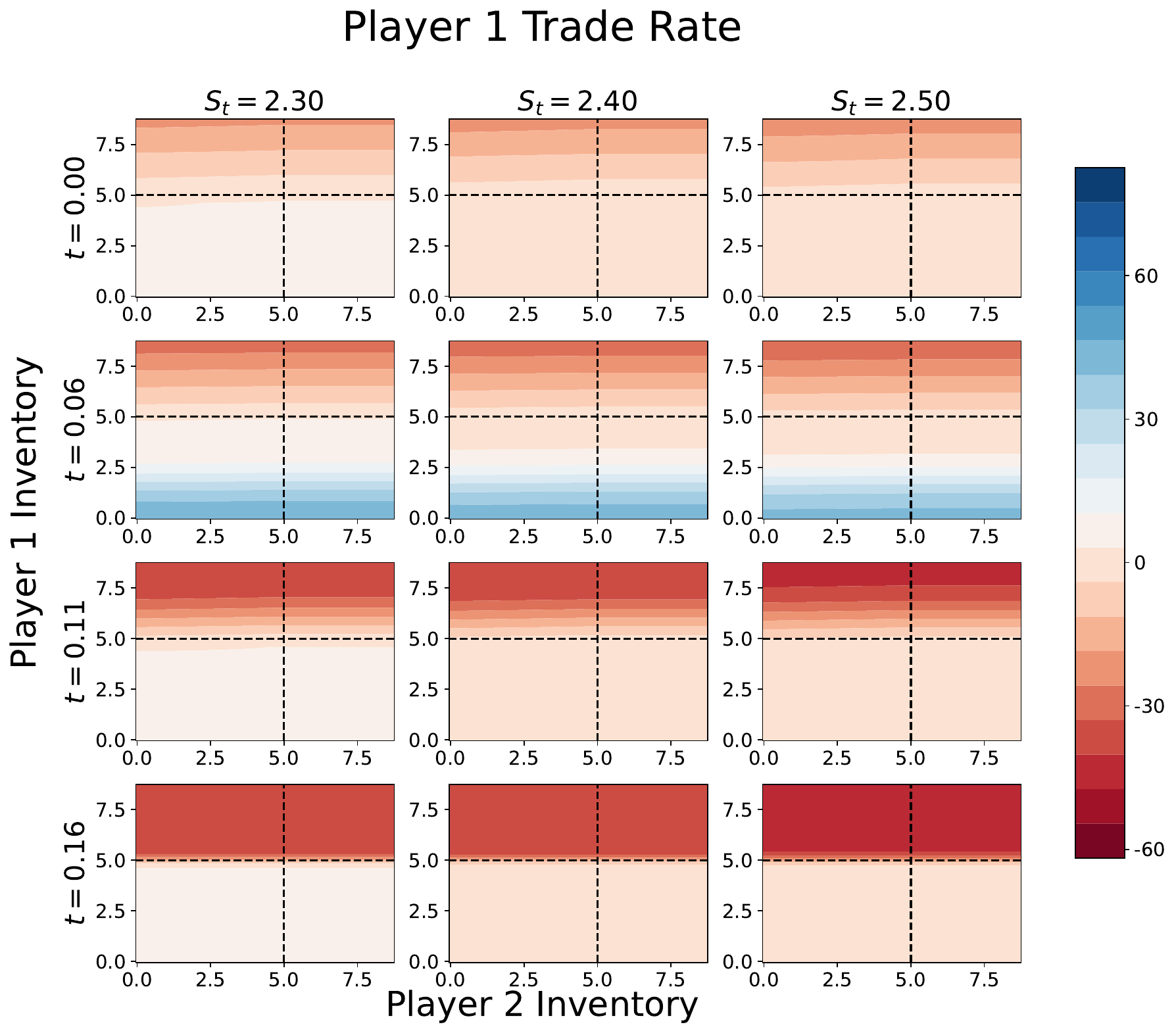} 
    \end{minipage}\vspace{0.25cm}
    \begin{minipage}{0.75\textwidth}
        \centering
        \includegraphics[width=\textwidth]{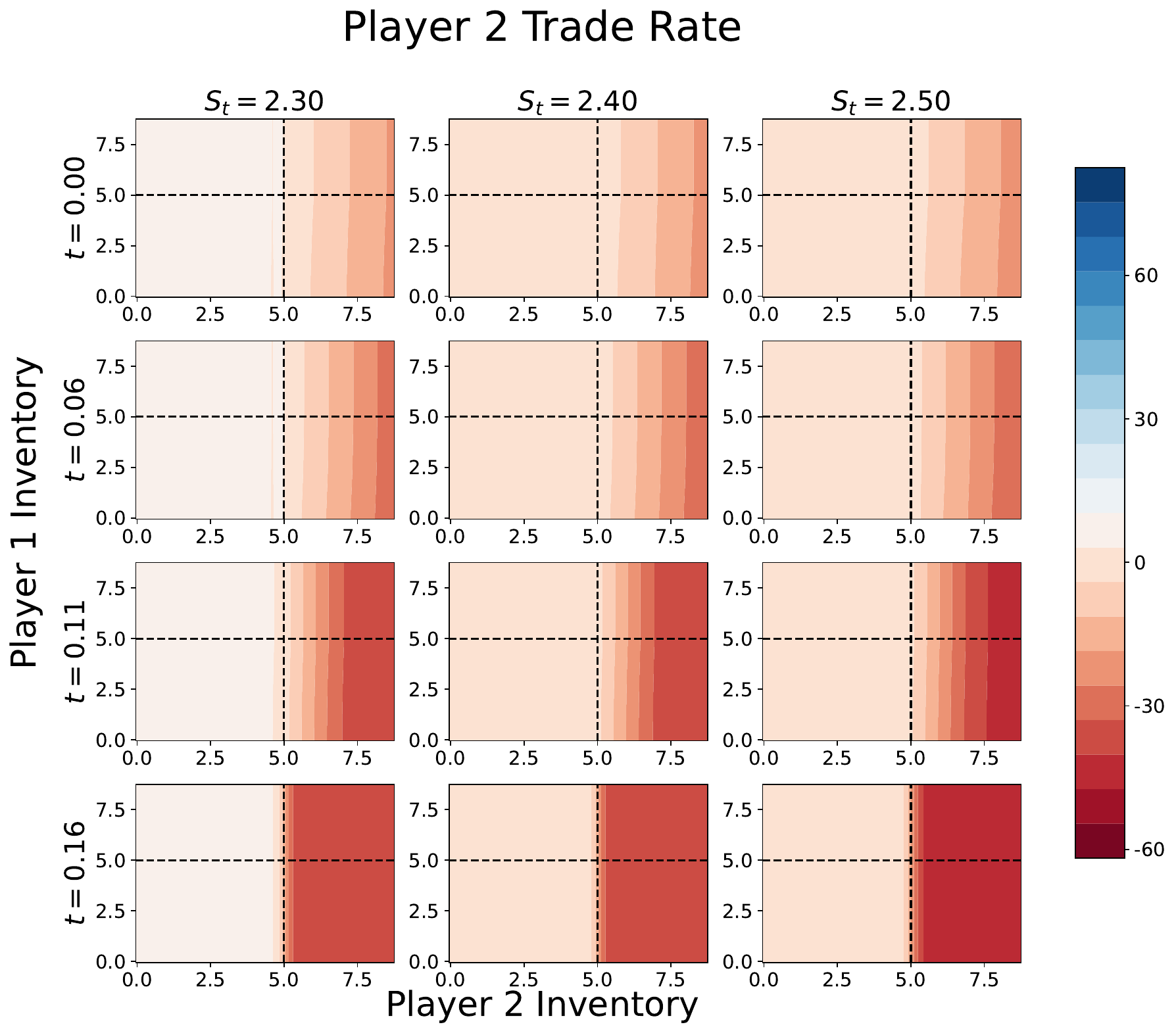} 
    \end{minipage}
    \caption{Trade rates conditional on players taking a trading action of players with heterogeneous generation capability in a two-period setting. The first two rows in each sub-figure correspond to the first period while the last two rows take place in the second period.}
    \label{fig:multi_trades}
\end{figure}

In Figures~\ref{fig:multi_probs} and~\ref{fig:multi_trades}, time points prior to $t=1/12$ are within the first compliance period, while time points after are within the second compliance period. Figure~\ref{fig:multi_probs} shows that as the first compliance date approaches in the first period, both players have large regions of high generation probability when below the OC requirement. {Both players' results are similar to those in the single-period setting in Section~\ref{sec:hetero} when analysing one period in the two-period setting.} The corresponding trade rates, shown in Figure~\ref{fig:multi_trades}, follow the same pattern as before, with positive trading and negative trading occurring when players are below and above their OC requirement, respectively. As there is more trading friction in this scenario, players are less inclined to trade at fast speeds and only do so when the gain is substantial. The boundaries between positive and negative trading become much more stark at the end of the final compliance period.

\begin{figure}[H]
\centering
\includegraphics[width=0.85\textwidth]{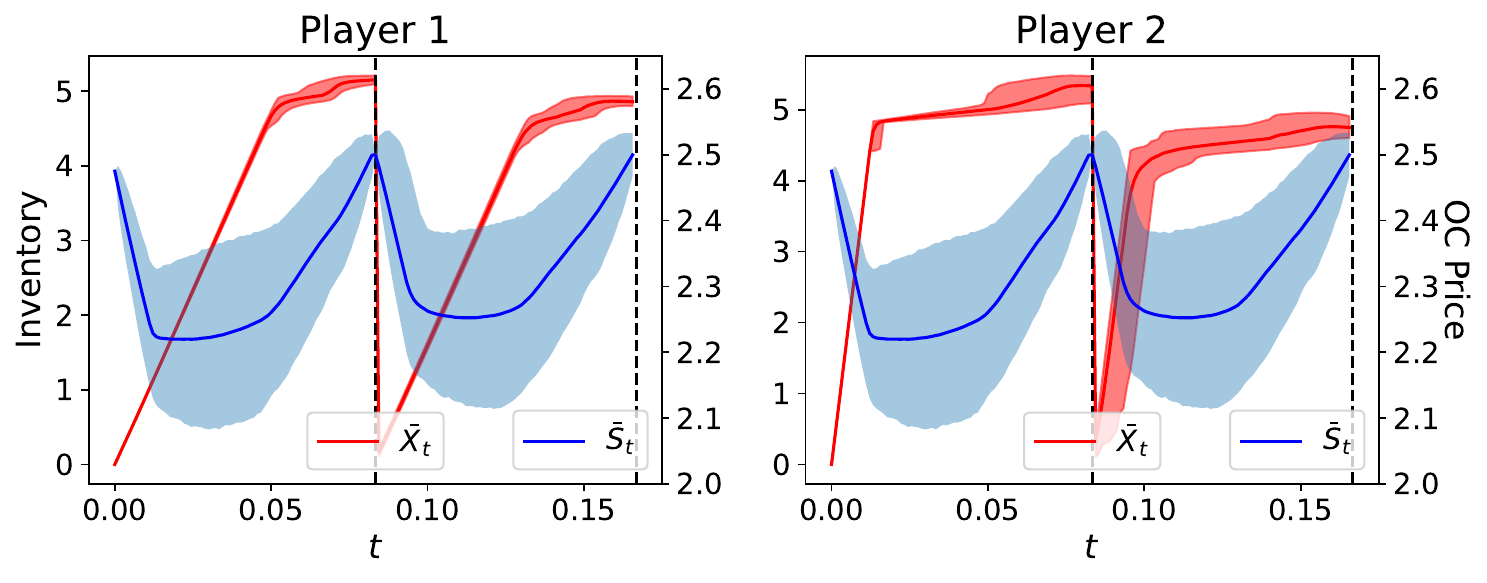}
\caption{Mean inventory with 95\% quantile (red) on left axis and mean OC price with 95\% quantile (blue) on right axis for heterogeneous players in multi-period model. Vertical dashed lines represent compliance dates.}
\label{fig:multi_inv}
\end{figure}

\begin{figure}[H]
\centering
\includegraphics[width=0.85\textwidth]{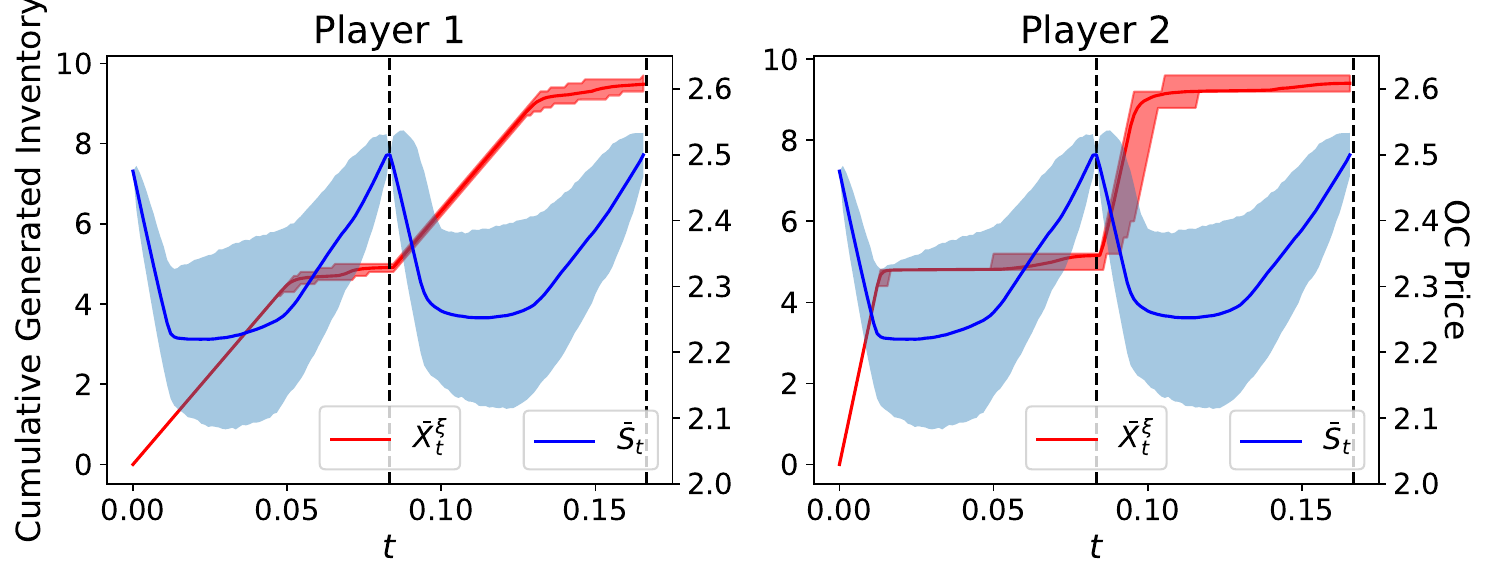}
\caption{Cumulative mean inventory acquired by generation with 95\% quantile (red) on left axis and mean OC price with 95\% quantile (blue) on right axis for heterogeneous players in multi-period model. Vertical dashed lines represent compliance dates.}
\label{fig:multi_inv_gen}
\end{figure}

Figures~\ref{fig:multi_inv} and~\ref{fig:multi_inv_gen} display the inventory and cumulative generated inventory for each player, respectively. In our simulation, both players slightly exceed the inventory requirement in the first period, as they are able to bank any excess OCs for the following period after compliance submission. When either player banks OCs, it provides them with more opportunities in the subsequent period to trade to help offset their costs. {Fewer generation instances occur in the second period, due to the previous periods banked OCs, resulting in a  moderate amount of increased variability in PnL, as demonstrated by player two's quantiles.}

The terminal PnL distribution, shown in Figure~\ref{fig:multi_hist_pnl}, clearly demonstrates that both players achieve beneficial outcomes than if they simply incurred the penalty in both compliance periods. The maximum penalty a player can incur after both periods $-\$25.00$. Players' terminal statistics, shown in Table~\ref{tab:multi_pnls} further demonstrate that both players attain non-trivial savings when abiding by the Nash equilibrium strategy despite the increased trading friction. Players eliminate all downside risk by a significant margin, as indicated by the players' {TE}.
\\[1em]
\begin{minipage}{0.45\textwidth}
\centering
\footnotesize
\begin{tabular}{ cccc } 
\toprule\toprule
  & & & PnL SE \\
  &  Mean PnL & {TE} & ($\times10^{-3}$) \\ 
 \midrule
 Player 1 & $-\$24.892$ &  $-\$24.950$ & 0.472 \\ 
 Player 2 & $-\$24.894$ & $-\$24.956$  & 0.477 \\ 
 \bottomrule\bottomrule
\end{tabular}
\captionof{figure}{Mean PnL, {TE}, and PnL SE for heterogeneous players in a two period model.}
\label{tab:multi_pnls}
\end{minipage}
\hfill
\begin{minipage}{0.45\textwidth}
\centering
\includegraphics[width=0.9\textwidth]{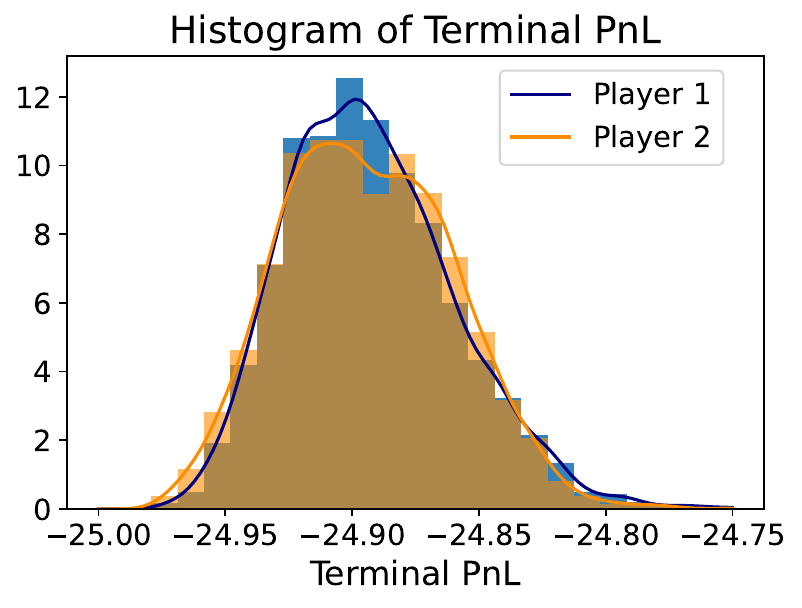}
\captionof{table}{Terminal PnL histograms of players with heterogeneous investment opportunities in a two period model.}
\label{fig:multi_hist_pnl}
\end{minipage}

\section{Conclusion}\label{sec:conc}

We developed a GHG OC model representative of Canada's market in a single-player and two-player setting. In the single-player setting, the player's value function solves a QVI, while in the two-player setting, players' value functions are determined by by computing the Nash equilibrium of a bi-matrix game. We found the scale of a player's OC generating projects plays a vital role in determining their behaviour and heavily influences the Nash equilibria of the two-player game. In the instances of the two-player game we analysed, players had well defined decision boundaries between trading OCs and generating OC. We find that players who {participate in the market with a heterogeneous counterpart perform better than those with a homogeneous counterpart. Here, better performance is measured with respect to terminal PnL and TE. This result stems from one of the two players in the heterogeneous setting have a larger generation capacity, despite having higher instantaneous investment cost.} The explicit-implicit FD scheme constructed for the numerical implementation allowed us to estimate the players' value functions and Nash equilibria avoiding numerical instabilities that would have been caused by the OC price's dynamics in an entirely explicit scheme. This choice of OC price dynamics allowed for a more realistic and arbitrage-free market model due to the convergence to the penalty value at the end of the compliance period. All simulations we performed demonstrated the importance of investing in GHG reducing and capturing projects from a financial perspective by completely mitigating downside risk, even under the strict choice that the marginal cost to generate one OC is equal to the penalty value. In more hostile trading environments, players still achieved financial gains when abiding by the equilibria. These financial incentives are in addition from the societal benefits these projects have on the earth's climate.

The market model and methodology presented in this paper will aid firms in deciding when to invest in projects that reduce or capture GHG emissions. In addition, this work allows legislators to more effectively organise OC and emissions markets by allow them to predict a market participant's behaviour. As nations strive for the goals set by the Paris Climate Accords and COP26, emissions markets and further GHG regulation are becoming more commonplace. Hence, the ability to properly model the resulting market is vital. Understanding player behaviour in GHG OC markets is vital to both the regulators, who can update the market framework given new findings, and the players, who must know how to optimally behave in order to offset their excess emissions. Both firms and the environment benefit from carbon capturing project implementation. As the world is reeling with the consequences of an increasingly warming and unstable climate, GHG OC markets and their regulation is one tool that can help slow the effects of climate change by creating incentives for firms to engage and invest in GHG emission reduction and capturing.

\subsection{Future Directions}

The model presented in this paper allows for multiple possible extensions, as both the analytical and numerical framework are adaptable to a number of problem instances. One natural extension is to characterise how players behave when presented with multiple different project opportunities. This can be done in both the single-player and two-player setting. In the single-player setting, this is accomplished by including additional terms to the player's QVI with multiple possible $\xi$ and $\cxi$ combinations. In the two-player setting, rows and columns would be added to the game matrices of the players to represent additional project opportunities for the players. 

A second worthwhile extension would be to pose this market model and optimisation problem in a reinforcement learning framework. One way to accomplish this is to employ the techniques developed by~\cite{casgrain2022deep}, which allows for the efficient computation of Nash equilibria in general-sum stochastic games. The flexibility of this RL framework would allow for easier scalability to a larger finite number of players that will better represent the future GHG OC market~(\cite{sadikman2022evolution}).

 \section{Acknowledgements}
 LW would like to thank Noah Marshall and Anthony Coache for helpful suggestions regarding numerical PDE methods. LW would like to acknowledge support from the Natural Sciences and Engineering Research Council of Canada via a Canadian Graduate Scholarship.
 SJ  would like to acknowledge support from the Natural Sciences and Engineering Research Council of Canada (RGPIN-2018-05705). Both LW and SJ would like to thank the associate editor and anonymous referee for the helpful and constructive comments.

\bibliographystyle{apalike}
\bibliography{refs.bib}

\appendix
    \section{Finite Difference Coefficients}\label{app:coefs}

    To solve for the coefficients of the FD scheme, we discretise the partial derivatives using central difference. Upon substituting the discretisations, the PDE~\eqref{eq:oneP_U_HJB} becomes
    \begin{equation}\label{eq:pin_dis_hjb}
        \begin{split}
            \frac{V_{k,i,j} -U^{\star}_{k-1,i,j}  }{\Delta t} + \frac{1}{2\kappa}\left(\frac{V_{k,i+1,j} -V_{k,i-1,j}  }{2\,\Delta x } - s_j \right)^2 &+ \frac{p-s_j}{T-t_{k-1}}\,\frac{U^{\star}_{k-1,i,j+1} -U^{\star}_{k-1,i,j-1}  }{2\,\Delta s } \\ &+ \frac{\sigma^2}{2}\,\frac{U^{\star}_{k-1,i,j+1} - 2U^{\star}_{k-1,i,j} +U^{\star}_{k-1,i,j-1} }{(\Delta s)^2} = 0\;.
        \end{split}
    \end{equation}
    By isolating for the terms $U^\star_{k-1,i,j-1},~U^\star_{k-1,i,j},~U^\star_{k-1,i,j+1}$, we construct the linear system (for a fixed $i$):
    \begin{align*}
        \frac{V_{k,i,j}}{\Delta t} &+ \frac{1}{2\kappa}\left(\frac{V_{k,i+1,j} -V_{k,i-1,j}  }{2\,\Delta x } - s_j \right)^2 =\\ &\frac{U^{\star}_{k-1,i,j}}{\Delta t} - \left(\frac{p-s_j}{T-t_{k-1}}\right)\frac{U^{\star}_{k-1,i,j+1} -U^{\star}_{k-1,i,j-1}  }{2\,\Delta s } -\frac{\sigma^2}{2}\frac{U^{\star}_{k-1,i,j+1} - 2U^{\star}_{k-1,i,j} +U^{\star}_{k-1,i,j-1} }{(\Delta s)^2}\;,\\
        \frac{H_{k,i,j}}{\Delta t} &= \left(\frac{p-s_j}{2\,\Delta s\,(T-t_{k-1})} - \frac{\sigma^2}{2\,(\Delta s)^2}\right)U^{\star}_{k-1,i,j-1} + \left(\frac{1}{\Delta t} + \frac{\sigma^2}{(\Delta s)^2}\right)U^{\star}_{k-1,i,j}\\& ~~~~~~~~-\left(\frac{p-s_j}{2\,\Delta s\,(T-t_{k-1})} + \frac{\sigma^2}{2\,(\Delta s)^2}\right)U^{\star}_{k-1,i,j+1}\;,\\
        H_{k,i,j} &= a_j \,U^{\star}_{k-1,i,j-1} + b_j \,U^{\star}_{k-1,i,j} +c_j \,U^{\star}_{k-1,i,j+1}\;.
    \end{align*}

\end{document}